# Information Theoretic One-Time Programs from Geometrically Local QNC$_0$ Adversaries

**Lev Stambler**[1]

Joint Center for Quantum Information and Computer Science, University of Maryland, College Park, MD 20742, USA

Department of Computer Science, University of Maryland, College Park, MD 20742, USA

NeverLocal Ltd. London, WC2H 9JQ, UK

March 28th, 2025

### Abstract

We show how to construct simulation secure one-time memories, and thus one-time programs, *without computational assumptions* in the presence of constraints on quantum hardware. Specifically, we build one-time memories from random linear codes and quantum random access codes (QRACs) when constrained to non-adaptive, constant depth, and $D$-dimensional geometrically-local quantum circuit for some constant $D$. We place no restrictions on the adversary's classical computational power, number of qubits it can use, or the coherence time of its qubits. Notably, our construction can still be secure even in the presence of *fault tolerant* quantum computation as long as the input qubits are encoded in a non-fault tolerant manner (e.g. encoded as high energy states in non-ideal hardware). Unfortunately though, our construction requires decoding random linear codes and thus does not run in polynomial time. We leave open the question of whether one can construct a *polynomial time* information theoretically secure one-time memory from geometrically local quantum circuits.

Of potentially independent interest, we develop a progress bound for information leakage via collision entropy (Renyi entropy of order 2) along with a few key technical lemmas for a "mutual information" for collision entropies. We also develop new bounds on how much information a specific $2 \mapsto 1$ QRAC can leak about its input, which may be of independent interest as well.

## 1 Introduction

One-time programs (and variants) were first formally introduced by Goldwasser et al. [1] and sit at the top of the cryptographic wish list. The primitive enables one-time proofs, one-time witness encryption, non-interactive secure two-party computation, and more [1]. Unfortunately, one-time programs are impossible to achieve in the standard model and idealized quantum model [1,2]. Still, if hardware assumptions enable a primitive known as one-time memories, then one-time programs can be built [1].

---

[1]levstamb@umd.edu

A one-time memory (OTM) is a cryptographic primitive which can be used to build one-time programs and non-interactive secure two party computation [3,4]. OTMs can be thought of as a non-interactive version of oblivious transfer (OT) where a sending party, Alice, sends a one-time memory to a receiving party, Bob. The memory encodes two classical strings, $s_0$ and $s_1$, and Bob can only learn one of the strings. After Bob reads $s_b$, any encoding of $s_{1-b}$ is "erased" and Bob cannot recover $s_{1-b}$.

Many OTM constructions rely on hardware-specific assumptions such as tamper-proof hardware [5,6] and trusted execution environments (TEEs) [7,8]. Broadbent et al. [2] also showed how to build (classical and quantum) one-time programs from OTMs in the Universal Composability (UC) model, while a line of works, [9–12], show how to use reusable hardware tokens to build OTMs. Ref. [13] also shows how to immunize one-time memories against quantum superposition attacks. In a line of works [14,15], Yi-Kai Liu shows how to build a weaker version of one-time memories in the isolated qubit model where a min-entropy bound is proved for the adversary. Liu then shows how to amplify security in the isolated qubit model to get secure single bit one-time memory [15]. Recently, interest in one-time programs has been renewed in the exploration of one-time programs for a restricted class of functionalities, mainly *randomized functionalities* [16,17]. Roehsner et al. [18] also show how to build probabilistic one-time programs using pairs of entangled qubits and experimentally demonstrate their construction.

## 1.1 Impossibility of One-Time Programs

To understand why hardware assumptions are necessary for one-time programs, we first need to understand why one-time programs are impossible in the standard model of quantum computation. For simplicity, we will consider deterministic one-time programs. Then, the impossibility follows from a simple observation: any deterministic functionality (or close to deterministic) can be reversed.

For example, after applying some circuit $U$ to a one-time program state with input $x$, an adversary can perform a weak measurement on some output register and then recover the original state of the program by applying $U^\dagger$. If the adversary can recover the original state, then the adversary can rerun the one-time program on a different input, $x'$. An important observation here is that the adversary must have some degree of *ideal hardware* to carry out this attack. For example, if the adversary's application of only $U^\dagger$ is noisy, then the adversary will not be able to recover the original program state!

## 1.2 Main Result

In this paper, our main result is a construction of one-time memories with statistical soundness in the presence of hardware constraints on quantum circuits. Specifically, we show that one-time memories are possible assuming that the adversary, given an input state, is constrained to applying a "geometrically-local and constant depth quantum circuit," which we term $\text{GQNC}_d^D$ for depth $d$ and geometric dimension $D$. Importantly, our adversary is only constrained in its *quantum computation* and can have unbounded classical computation. Thus, the adversary can *internally simulate any quantum computation* using its classical resources. We believe that this adversarial setting is reasonable when considering a world of heterogeneous quantum hardware. For example, consider a world with wide-spread quantum internet where the underlying qubits are quite noisy and non-amenable to error-corrected computation. Yet, in this setting, fault tolerant quantum computation may still be possible using different hardware.

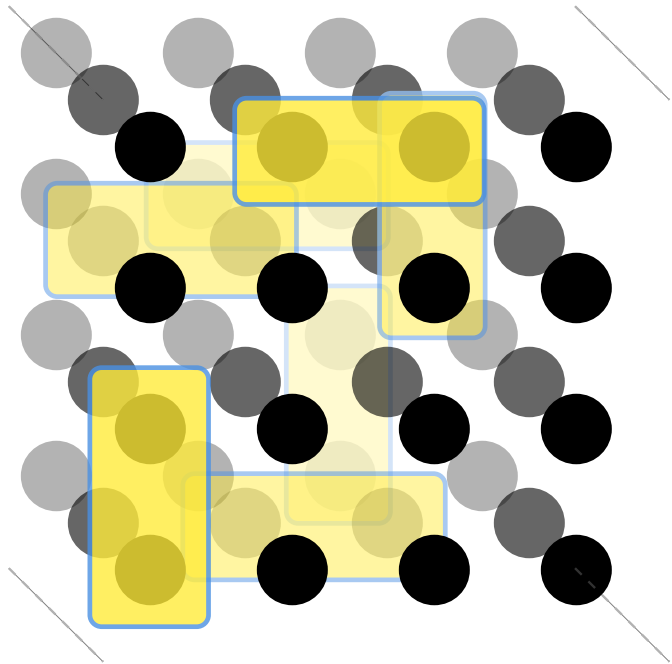

Figure 1: An example of a $\text{GQNC}_3^2$ circuit where we have a depth 3 circuit on a 2D grid of qubits. Each "layer" of the image represents a different time step in the circuit and each black dot represents a qubit within each layer. The yellow boxes represent the gates acting on a pair of qubits.

Specifically, we note that given existing QKD infrastructure, we justify two of the three assumptions of the $\text{GQNC}_d^D$ model as follows:

- Even though non-local hardware is ideally supposed to have no geometric-locality constraints, there is still a bound on locality when considering limits on the depth of computation and decoherence when moving qubits. So, we believe that these restrictions can be captured by a *geometrically-local restriction* for some fixed dimension $D$. We do not believe that the geometric restriction is fundamental to our result and hope to remove it in future work as well.

- The *depth bound* on quantum circuits is justified by the experimental evidence of noisy gates. So, beyond a certain fixed depth, the noise in the circuit will be too high to be useful.

Lastly, though recent works have shown that mid-circuit measurements can be performed in a broad variety of hardware [19–21], we believe that our reliance on the non-adaptive nature of the adversary can be removed in future work with a more careful analysis of the adversary's quantum circuit.

Succinctly, we get the following result:

> *Theorem 1.1* (One-time Memory from Constrained Adversaries (Informal)): Assuming that the adversary is constrained to $C^{\text{1-GQNC}_d^D}$, where $C$ represents unconstrained classical computation and 1-GQNC$_d^D$ represents oracle access to a single application of a GQNC$_d^D$ circuit, we can construct a one-time memory with statistical security and negligible correctness error.

Unfortunately, our presented algorithm requires exponential running time thus making it impractical. However, we believe that our result is a step in the right direction and can be improved in future work.

**Related Work**

Most similar to our work, Qipeng Liu shows how to construct one-time memory in the NISQ regime assuming a bound on coherence time and time-lock puzzles [22]. In particular, Liu uses conjugate coding to encode two messages, one in the computational basis and one in the Hadamard basis. Then, the receiving party chooses a basis to measure in and solves a time-lock puzzle to learn which qubits contain the message for the chosen basis. The assumption on the adversary's coherence time is crucial for the security of the protocol.

In contrast, our work makes no computational/ oracular assumption (outside the limits on the adversary's quantum circuit) and no assumption on the coherence time of the adversary. On the other hand, our work does assume that the adversary's quantum circuit is non-adaptive and geometrically constrained unlike Liu's work. Moreover, our construction does not run in polynomial time.

Yi-Kai Liu also shows how to build variations of one-time memories in the isolated qubit model without computational assumptions [14,15,23]. The key idea is to encode two messages into Wiesner states and then use near capacity error correction to recover one of the messages. Similar to our work, Liu relies on the codes being close to capacity and providing a sort of "k-wise" independence for each set of isolated qubits. On the contrary, we use random access codes instead of Wiesner states to show a progress bound on the adversary's information gain. Moreover, Liu's work provides min-entropy bounds on the adversary's knowledge of both messages but not simulation-based security for large message spaces. Again though, Liu's work does not make a fixed quantum circuit assumption, while our work does.

### 1.3 Technical Overview

In this technical overview, we outline the main techniques used in our construction of one-time memories from $C^{\text{1-GQNC}_d^D}$ adversaries. In this exposition, we will focus on one-time memories for random messages, with one-time memories for all messages following from standard techniques in Section 7.

The main technique is to combine binary linear error correction [24] with $2 \mapsto 1$ quantum random access codes (QRAC) [25]. A $2 \mapsto 1$ QRAC[2] is a quantum protocol where a sender, Alice, can encode two bits $b_0, b_1$ into a single qubit state $\mathcal{E}(b_0, b_1) \in \mathbb{C}^2$. Then, the receiver, Bob, can perform measurement $\mu_\alpha$, for $\alpha \in \{0,1\}$, on $\mathcal{E}(b_0, b_1)$ to recover $b_\alpha$ with some maximum

---
[2]Throughout the paper, we use the specific and optimal $2 \mapsto 1$ QRAC of Ref. [25].

probability of error. Specifically, the sender, Alice, encodes two random messages $m_0, m_1$ as follows:

1. Let $\mathcal{C}_0, \mathcal{C}_1 \subset \mathbb{F}_2^n$ be two publically known and fixed random linear codes and $c_0 \in \mathcal{C}_0, c_1 \in \mathcal{C}_1$ be two codewords corresponding to $m_0$ and $m_1$ respectively.
2. Let $\mathcal{E}(c_0^i, c_1^i)$ be a $2 \mapsto 1$ QRAC state for each $i \in [n]$.
3. Alice then sends the QRAC states to Bob.

To recover $m_\alpha$, Bob performs the following steps:

1. Measure each qubit, $\mathcal{E}(c_0^i, c_1^i)$, using $\mu_\alpha$ to recover noisy $c'_\alpha$ with some probability of error.
2. Use the error correction property of $\mathcal{C}_\alpha$ to recover $m_\alpha$ from the measurements.

It is not too hard to see that the above protocol is correct assuming that the coding parameters are properly chosen. Note that Bob recovers $m_\alpha$ as long as $c'_\alpha$ is close to $c_\alpha$ in hamming distance. This is guaranteed by the properties of the quantum random access code.

Soundness, however, is more difficult to show. To do so, we have to build up some technical machinery. First, we develop some natural extensions of collision entropy and mutual information. Then, we use a combination of computational and analytic methods to show some facts about single-qubit $2 \mapsto 1$ quantum random access codes. Finally, we use these facts to show a progress bound on the adversary's information gain in the presence of a $\text{GQNC}_d^D$ circuit.

**Collision Entropy and Mutual Information**

As often done in cryptography, we adopt the notion of *collision entropy* and extend its definition to encompass a form of mutual information [26–28]. Unlike Shannon entropy, we can use collision entropy in a cryptographic setting due to a simple application of Markov's inequality. And, unlike minimum entropy, collision entropy is not a measure of the worst-case scenario but rather a measure of the average case scenario which can then be adopted to the worst-case scenario via a Markov bound.

In this work, we also define collision-mutual information as a measure of the "information gain" of an adversary when considering the collision entropy of two random variables. Specifically, it is simply a measure of the difference between the collision entropy of a random variables when conditioned on another random variable.

We define our version of entropy in a similar manner to how Refs. [29,30] define Renyi entropy[3]:

> *Definition 1.1* (Collision Information Theory (informal)): We use the same definition as the standard collision entropy though with a slightly different moral connotation
> - $\mathsf{H}_c(X) = -\log_2 \mathbb{E}_x[\Pr[X = x]]$
> - $\mathsf{H}_c(X \mid Y) = -\log_2 \mathbb{E}_{x,y}[\Pr[X = x \mid Y = y]]$
> - $\mathsf{I}_c(X : Y) = \mathsf{H}_c(X) - \mathsf{H}_c(X \mid Y)$

We use many standard properties of collision entropy and its mutual information equivalent culminating in the following key lemma, reminiscent of how conditional Renyi entropy is used in [31]:

---

[3]We note that collision entropy is equivalent to Renyi entropy of order 2

**Fact 1.1** (Collision Entropy Chain Rule (informal)): Let $X, Y, Z$ be random variables such that
$$\mathsf{I}_c(X:Z) \leq s - B$$
alongside the following condition:
$$\mathsf{I}_c(X:Y \mid Z) = \mathsf{H}_c(X \mid Z) - \mathsf{H}_c(X \mid YZ) \leq C.$$
*if* $H_{\min}(X \mid Z) \geq \mathsf{H}_c(X) - s$. Then,
$$\mathsf{I}_c(X:Y \mid Z) \leq C$$
for a statistically close distribution to the original distribution.

**Facts About $2 \mapsto 1$ Quantum Random Access Codes**

Next, we develop key bounds on QRAC codes with uniform and independent input bits. Specifically, we will want to bound how much a $\mathrm{GQNC}_d^D$ circuit can learn about the input bits $b_0, b_1$ after running the QRAC protocol.

We will be specifically interested in the following three quantities: (1) the greater bit mutual information ($I_>$), (2) the total bit mutual information ($I_{\mathrm{tot}}$), and (3) the conditional bit mutual information ($\tilde{I}_>$). The greater bit mutual information can be thought of as the maximum of the mutual information of each input bit with the output state. Formally, for two IID bit strings $b_0, b_1 \in \{0,1\}^m$, we want to bound

$$\sup_\mu \max_{y \in \{0,1\}^m} \mathsf{I}_c \left[ b_{y_1}^1, ... b_{y_m}^m : \mu \left( \bigotimes_{i \in [m]} \mathcal{E}(b_0^i, b_1^i) \right) \right] \leq I_> \cdot m$$

for constant $I_>$ and all possible measurements $\mu$. Then, we can think of the total bit mutual information as the sum of the mutual information of each input bit with the output state. We want to bound

$$\sup_\mu \mathsf{I}_c \left[ b_0^1, ... b_0^m : \mu \left( \bigotimes_{i \in [m]} \mathcal{E}(b_0^i, b_1^i) \right) \right] + \mathsf{I}_c \left[ b_1^1, ... b_1^m : \mu \left( \bigotimes_{i \in [m]} \mathcal{E}(b_0^i, b_1^i) \right) \right] \leq I_{\mathrm{tot}} \cdot m$$

for constant $I_{\mathrm{tot}}$. Finally, we will want a similar bound for the *conditional bit mutual information* where we condition on one of the input bits. Formally, we want to bound, for IID string $b_\alpha$ and arbitrary string $b_{1-\alpha}$,

$$\sup_\mu \mathsf{I}_c \left[ b_\alpha^1, ..., b_\alpha^m : \mu \left( \otimes_{i \in [m]} \mathcal{E}(b_0^i, b_1^i) \right) \mid b_{1-\alpha} \right] \leq \tilde{I}_> \cdot m$$

for constant $\tilde{I}_>$.

We use a combination of computational techniques for single qubits and analytical techniques to get the following bounds:

**Fact 1.2**: Given the optimal $2 \mapsto 1$ QRAC scheme in [25], we get that
$$I_> \leq 0.59 \qquad I_{\mathrm{tot}} \leq 0.65 \qquad \tilde{I}_> \leq 0.59$$

**OTMs for Random Messages**

We can now combine the above lemmas to show soundness for the outlined protocol. Specifically, we must show that the collision entropy of either $c_0$ or $c_1$ is sufficiently high after the adversary's run of the $\text{GQNC}_d^D$ circuit. In the following exposition, assume that the adversary learns more information about $c_0$ than $c_1$. We first make use of our restrictions on the adversary in the following way:

1. The *non-adaptive* property of the quantum circuits is used to "reorder" the order of measurements which the adversary makes. Thus, if we partition the input qubits into different sets, we can rearrange the adversary's order of measurements such that all measurements in the reverse light-cone of a set of qubits are made consecutively. We can then use this reordering to get a progress bound on the amount of information an adversary learns with each set of measurements.

2. The *geometrical locality* and *depth bound* are used to create two sets of input qubits: "shell" qubits and "light-cone independent" qubits. The "light-cone independent" input qubits are further partitioned into sets of qubits where each set shares an *independent* reverse light-cone from the other sets. Because of the geometric-locality and depth bound, the cardinality of the independent sets are all upper-bounded by some constant. The "shell" qubits do not have any nice properties which allow us to bound how much an adversary learns from them. So, in the soundness argument, we can pretend to give the adversary all the information in the "shell qubits." As long as the proportion of shell is small, we can show that soundness still holds even if the adversary learns all the information in the shell qubits.

In more detail, let the partition of light-cone independent qubits be the sets $\mathtt{cu}_1, ... \mathtt{cu}_q$.[4] We will let $\mu_1, ..., \mu_q$ be the measurement results of the adversary on the qubits in $\mathtt{cu}_1, ..., \mathtt{cu}_q$ respectively. Due to the non-adaptivity, we will assume that the adversary first measures the qubits in $\mathtt{cu}_1$, then $\mathtt{cu}_2$, and so on. Also, note that for $\alpha \in \{0, 1\}$, $\mathsf{H}_c(c_\alpha) = k$ where $k$ is the dimension of the code $\mathcal{C}_\alpha$. So, if we assume that, in the worst-case, the adversary learns all the information in the shell qubits, denoted as $\overline{\mathtt{CU}}$, we have that $\mathsf{H}_c(c_\alpha \mid \overline{\mathtt{CU}}) \geq k - O(|\overline{\mathtt{CU}}|)$. Given that we can assume that $\frac{|\overline{\mathtt{CU}}|}{n} \to 0$ as $n$ grows, we can show that the shell qubits do not reveal too much information to the adversary.

Next, recall that $c_\alpha = G_\alpha \cdot m_\alpha$ for some $m_\alpha \in \{0,1\}^k$ and that $G_\alpha$ is a random $n \times k$ matrix. Then, we can see that the qubits in $\mathtt{cu}_1$ depend only on classical information generated by $G_0[\mathtt{cu}_0] \cdot m_0$ and $G_1[\mathtt{cu}_1] \cdot m_1$. If we set $|\mathtt{cu}_0|$ to be small enough, we then get that $G_\alpha[\mathtt{cu}_\alpha]$ is a 2-universal hash function. So, assuming that $\mathsf{H}_c(c_\alpha \mid \overline{\mathtt{CU}}) \gg |\mathtt{cu}_0|$, we can make use of the left-over hash lemma[5] in conjunction with Fact 1.1 (the correspondence between $\mathsf{H}_c$ and $H_{\min}$) to show that $c_0[\mathtt{cu}_0]$ and $c_1[\mathtt{cu}_0]$ are indistinguishable from random. Thus, for all $i \in \mathtt{cu}_0$, $\mathcal{E}(c_0^i, c_1^i)$ is indistinguishable from an independent and random QRAC state!

Now that we have that $\mathcal{E}(c_0^i, c_1^i)$ is indistinguishable from random, we can bound how much an adversary can learn about $c_0$ and $c_1$ from the qubits in $\mathtt{cu}_1$ by $|\mathtt{cu}_1| \cdot I_{\text{tot}}$ and how much the adversary can learn about $c_0$ or $c_1$ individually by $|\mathtt{cu}_1| \cdot I_>$ using Fact 1.2. We then get the following lower bound: $\mathsf{H}_c(c_\alpha \mid \overline{\mathtt{CU}}, \mu_1) \gg |\mathtt{cu}_2|$. We repeat the process for $\mathtt{cu}_2$, showing that the qubits within $\mathtt{cu}_2$ are indistinguishable from uniformly random and independent QRAC states and thus get a bound $\mathsf{H}_c(c_\alpha \mid \overline{\mathtt{CU}}, \mu_1, \mu_2) \gg |\mathtt{cu}_3|$ for both $\alpha \in \{0, 1\}$. We continue this process inductively until the adversary learns most of the information for $c_0$. We then bound the remaining information which the adversary can learn about $c_1$ in a similar manner. The only difference is that now only $c_1$ has a significant amount of entropy. We thus use the conditional bit mutual information to show that the adversary cannot learn more that $\tilde{I}_> \cdot |\mathtt{cu}_j|$ bits of information about $c_1$ for measurements on $\mathtt{cu}_j$.

---

[4] $\mathtt{cu}$ stands for (hyper)cube.

[5] Recall that the left-over hash lemma states that if $H_{\min}(X) \geq m + \lambda$, then for a public two-universal hash function, $h$, $h(X)$ is $2^{\lambda/2}$ close to random.

Finally, we can show that after all measurements are made, $H_c(c_1 \mid \overline{\mathtt{CU}}, \mu_1, ..., \mu_q) \gg \lambda$ for some security parameter $\lambda$. Thus, the adversary cannot learn much about $c_1$ and so the protocol is sound.

**Notation and Organization**

Throughout this paper, we will use several notational conventions. Generally, we will use lowercase letters for vectors and uppercase letters for matrices. When referring to the elements of a vector or matrix, we will use superscript notation, e.g. $x^i$ refers to the $i$th element of vector $x$ and $A^i$ to refer to the $i$-th row of matrix $A$. For sets, we will generally use script letters such as $\mathcal{S}$, and for any set $\mathcal{S}$, $|\mathcal{S}|$ will denote its cardinality. We will also abuse notation slightly such that $s[\mathcal{S}]$ denotes the bits indexed by the set $\mathcal{S}$. Similarly, we will write $A[\mathcal{S}]$ to denote the submatrix formed by the rows of $A$ indexed by the set $\mathcal{S}$. Finally, for any function $f$, $f(\cdot)$ represents its evaluation.

The paper is organized as follows. In Section 2, we provide a brief overview of the necessary background material for the rest of the paper. In Section 3, we introduce the notion of collision entropy and its smooth variants. We also provide a few key technical lemmas for a sort of "mutual information" for collision entropies. Then, in Section 4, we define the greater and total bit mutual information for a $2 \mapsto 1$ QRAC scheme and show how to calculate them. In Section 5, we formally define the complexity class $\mathrm{GQNC}_d^D$ and the adversary class $C^{1\text{-}\mathrm{GQNC}_d^D}$. Next, in Section 6, we provide our main technical result: a construction of unconditionally sound one-time memories for random messages. In Section 7, we show how to build one-time memories from one-time random memories. Finally, in Section 8, we conclude with a discussion of our results and potential future directions.

# 2 Background

In this section, we provide a brief overview of the necessary background material for the rest of the paper.

## 2.1 Quantum Random Access Code (QRAC)

A quantum random access code is a uniquely quantum primitive that allows a sender to encode multiple messages into a single quantum state. A receiver can then perform a measurement on the state to recover one of the messages with some probability of error.

*Definition 2.1* ($2 \mapsto 1$ Quantum Random Access Code (QRAC) [[25]]): A QRAC is an ordered tuple $(E, D_0, D_1)$ where $E : F_2^2 \to \mathbb{C}^2$ and 2 sets of orthogonal measurements $M_i = \{|\varphi_\alpha^0\rangle, |\varphi_\alpha^1\rangle\}$ for $\alpha \in \{0, 1\}$.

For our purposes, we will consider a fixed $2 \mapsto 1$ QRAC. Specifically, we will use the optimal $2 \mapsto 1$ QRAC from [25].

Define $|\psi_\theta\rangle = \cos(\theta)|0\rangle + \sin(\theta)|1\rangle$.

*Definition 2.2* (Canonical and optimal $2 \mapsto 1$ QRAC): Then, the optimal $2 \mapsto 1$ QRAC is given by the following measurements:
- $\mathcal{E}(0,0)$ returns $|\psi_{\frac{\pi}{8}}\rangle$
- $\mathcal{E}(0,1)$ returns $|\psi_{-\frac{\pi}{8}}\rangle$
- $\mathcal{E}(1,0)$ returns $|\psi_{5\frac{\pi}{8}}\rangle$
- $\mathcal{E}(1,1)$ returns $|\psi_{-5\frac{\pi}{8}}\rangle$

and the measurements are given by:
- $\mu^0 = \{|0\rangle, |1\rangle\}$ (the $Z$ basis)
- $\mu^1 = \{|\psi_{\frac{\pi}{4}}\rangle, |\psi_{-\frac{\pi}{4}}\rangle\}$ (the $X$ basis)

We also know that the optimal success probability of this QRAC is $\cos^2\left(\frac{\pi}{8}\right) \approx 0.85$.

## 2.2 Classical Error Correcting Codes

A classical error correcting code is a mapping from a message to a codeword such that the codeword can be recovered from a noisy version of the codeword. A linear code is a code where the codewords form a linear subspace of the vector space. Specifically, we will use the following definition.

*Definition 2.3* (Linear Code, [24]): A linear code, $\mathcal{C}$ is a code defined by its generator matrix $G \in \{0,1\}^{n \times k}$ and parity check matrix $H \in \{0,1\}^{n \times (n-k)}$ such that
$$\mathcal{C} = \{x \in \{0,1\}^n : Hx^T = 0\}$$
which is equivalent to
$$\mathcal{C} = \{x \in \{0,1\}^n : x = Gy^T \text{ for some } y \in \{0,1\}^k\}$$

We also define $\text{ECDec}(\mathcal{C}, y)$ to be the error correcting decoding algorithm for code $\mathcal{C}$ and received word $y$ which maps $y$ to the closest codeword in $\mathcal{C}$.

Moreover, the rate of a linear code is defined as
$$R = \frac{\log_2(|\mathcal{C}|)}{n} = \frac{k}{n}$$

We say that a code can correct errors with probability $\varepsilon_{\text{corr}}$, from a binary symmetric channel with error rate $p$ if for $y = c + e$ where $e$ is a noise vector and $c$ is a codeword, we have that
$$\Pr_{[e,c]}[\text{ECDec}(\mathcal{C}, y) \neq c] \leq \varepsilon_{\text{corr}}$$

## 2.3 Min Entropy, Statistical Distance, and Smooth Min Entropy

Smooth notions of entropy are useful in the context of quantum information theory and cryptography as they allow us to consider the entropy of a random variable that is close to another random variable in statistical distance.

*Definition 2.4* (Min-Entropy): Minimum binary entropy of a random variable $X$ is defined as
$$H_{\min}(X) = -\log_2\left(\max_x \Pr[X = x]\right)$$

*Definition 2.5* (Statistical Distance): For two probability distributions $p, q$, the statistical distance is defined as

$$\mathrm{SD}(p,q) = \frac{1}{2}\sum_x |p(x) - q(x)|$$

*Definition 2.6* (Smooth Min Entropy [32–34]): The smooth minimum binary entropy of a random variable $X$ is defined as
$$H_{\min}^{\epsilon(X)} = \max_{X'}\bigl(H_{\min(X')}\bigr)$$
where $\mathrm{SD}(X, X') \leq \epsilon$.

## 2.4 Left Over Hash Lemma and Randomness Extractors

A key part in the upcoming proof is the celebrated Left Over Hash Lemma which states that if a random variable has min entropy greater than a certain threshold, then it is close to uniform in statistical distance after applying a 2-universal hash function.

**Lemma 2.1** (Left Over Hash Lemma [35]): For a random variable $X$ with min-entropy $H_{\min}(X)$, and a 2-universal hash function $h : \mathcal{S} \times \mathcal{X} \to F_2^m$, we have that if
$$H_{\min}(X) \geq m + 2\log\left(\frac{1}{\epsilon}\right)$$
then
$$\mathrm{SD}[(h(S,X), S), (U, S)] \leq \epsilon$$
for $U$ uniform on $F_2^m$ and randomly sampled $S$.

*Definition 2.7* (Statistical Randomness Extractor): A $(k, \epsilon)$-randomness extractor is a function $\mathtt{Ext} : \mathcal{S} \times \mathcal{X} \to F_2^m$ such that for any $X$ with $H_{\min}(X) \geq k + O\bigl(\log(\frac{1}{\epsilon})\bigr)$,
$$\mathrm{SD}[(S, \mathtt{Ext}(S,X)), (S, U)] \leq \epsilon$$
for $U$ uniform on $F_2^m$ and randomly sampled $S$.

**Lemma 2.2** (Randomness Extractor Parameters [36]): For $k \in [0, n], \epsilon > 0$, there exists an $(k, \epsilon)$-randomness extractor $\mathrm{extract} : F_2^n \times F_2^d \to F_2^m$ if $H_{\min}(X) \geq k$, $d = \log(n - k) + 2\log(\frac{1}{\epsilon}) + O(1)$, and $m = k + d - 2\log(\frac{1}{\epsilon}) - O(1)$.

Notice that we can set $d \leq \log(n) + 2\log(\frac{1}{\epsilon}) + O(1)$ and $k \geq m + 2\log(\frac{1}{\epsilon}) + O(1) - \bigl(\log(n-k) + 2\log(\frac{1}{\epsilon}) + O(1)\bigr)$ and so we can set $k \geq m + O(1)$. We will primarily use this parameter setting in the construction of the one-time memory.

## 2.5 One Time Memories

Though more complex definitions of one-time memories exist (such as the UC-based definitions in [2]), we will use a simpler definition which is very similar to [22] for the purposes of this paper. Because we do not assume any assumptions/heuristics beyond the restrictions on the quantum adversary, we can simplify the soundness property from [22] to be non-oracular. We also strengthen the soundness to be sub-exponential in the security parameter rather than super-polynomial.

*Definition 2.8* (One-Time Memory): A one-time memory is a protocol between a sender and receiver which can be represented as a tuple of algorithms (prepState, readState) where:
- prepState is a probabilistic algorithm which takes as input $s_0, s_1 \in \mathcal{S}$ and outputs a quantum state $\rho$ as well as classical auxiliary information aux.
- readState is a (potentially probabilistic) algorithm which takes as input $\rho$, aux and $\alpha \in \{0, 1\}$ and outputs a classical string $s_\alpha$ with probability $1 - \epsilon_{\text{corr}}$.

*Definition 2.9* (Correctness): A one-time memory (prepState, readState) is said correct with probability $\epsilon_{\text{corr}}$ if for all $s_0, s_1 \in \mathcal{S}$, we have that
$$\Pr[s_\alpha = s_{\alpha'}] \geq 1 - \epsilon_{\text{corr}}$$
where $s_\alpha = \text{readState}(\text{prepState}(s_0, s_1), \alpha)$.

*Definition 2.10* (Soundness): A one-time memory (prepState, readState) is said to be sound relative to an adversary, $\mathcal{A}$, which interacts with the protocol, if there exists a simulator Sim for every inverse sub-exponential $\gamma(\cdot)$ for every $s_0, s_1 \in \mathcal{S}$ such that Sim makes at most one query to $g^{s_0,s_1} : \{0, 1\} \to \{s_0, s_1\}$ (where $g(\alpha) = s_\alpha$) and
$$\mathcal{A}\big(\text{prepState}(1^\lambda, s_0, s_1)\big) \stackrel{\gamma(\lambda)}{\approx} \text{Sim}^{g^{s_0,s_1}}(1^\lambda)$$
where $\stackrel{\gamma(\lambda)}{\approx}$ denotes statistical distance of at most $\gamma(\lambda)$.

## 3 Collision Entropy and Information Theory

Shannon's Information theory is a very useful tool when attempting to quantify the degree of uncertainty. Unfortunately, Shannon theory can be quite challenging to apply to cryptographic protocols which require **worst-case** security guarantees. On the other hand, cryptographic notions of entropy, such as min-entropy, though suitable for worst-case security, can often be "too pessimistic." For example, consider a probabilistic process which outputs secrets $s_i \in \{0, 1\}^2$ with probability $\frac{1}{2}$ and otherwise outputs random $r_i$ Then, for $n$ secrets, the min-entropy of the output is $n$. But, the secrecey of $s_1, ..., s_n$ may be more than $n$ bits as an adversary may receive $s_i$ with probability $\frac{1}{2}$ but also **does not know** if the output of the process is $s_i$ or $r_i$. Thus, there is some extra information which min-entropy does not capture.

In the context of this work, we will use collision-based notions of entropy (also known as Renyi Entropy of order 2). For a more detailed comparison of Shannon and Renyi entropy, see [28]. Missing proofs for this section can be found in Appendix A.

Though multiple notions of conditional Renyi entropy exist, we will use the following definitions.

*Definition 3.1* (Collision Information): The collision entropy for random variable $X$ is defined as follows
$$\mathsf{H}_c(X) = -\log_2 \mathbb{E}_x[\Pr[X = x]] = -\log_2 \left( \sum_x \Pr[X = x]^2 \right)$$
and conditional collision entropy as

$$\mathsf{H}_c(X \mid Y) = -\log_2\big(\mathbb{E}_{x,y}[\Pr[X = x \mid Y = y]]\big)$$
$$= -\log_2 \mathbb{E}_y \mathbb{E}_x[\Pr[X = x \mid Y = y]]$$
$$= -\log_2\left(\sum_{x,y} \Pr[X = x, Y = y] \cdot \Pr[X = x \mid Y = y]\right).$$

We then define the collision mutual information as
$$\mathsf{I}_c(X : Y) = \mathsf{H}_c(X) - \mathsf{H}_c(X \mid Y) = \log_2\left(\frac{\mathbb{E}_{x,y}[\Pr[X = x \mid Y = y]]}{\mathbb{E}_x[\Pr[X = x]]}\right)$$

and the conditional collision mutual information as
$$\mathsf{I}_c(X : Y \mid Z) = \mathsf{H}_c(X \mid Z) - \mathsf{H}_c(X \mid Y, Z) = \log_2\left(\frac{\mathbb{E}_{x,y,z}[\Pr[X = x \mid Y = y, Z = z]]}{\mathbb{E}_x[\Pr[X = x \mid Z = z]]}\right)$$

We will also define $\mathsf{H}_{c_p}(X \mid Y), \mathsf{I}_{c_p}(X : Y \mid Z)$ as the collision entropy and collision mutual information for distribution $p$ over $X, Y, Z$.

Just as with Shannon/ worst-case notions of entropy, we can define smooth collision notions entropy.

*Definition 3.2* (Smooth Collision Entropy): For a random variable $X$ with probability distribution $p$, we define the smooth collision entropy as
$$\mathsf{H}_c^\varepsilon(X \mid Y) = \max_{X', Y' \sim X, Y} \mathsf{H}_c(X' \mid Y')$$
where the maximum is over all $(X', Y')$ such that $\mathrm{SD}((X,Y),(X',Y')) \leq \varepsilon$.

We similarly define "smooth" collision mutual information.

*Definition 3.3* (Smooth Collision Mutual Information): For random variables $X, Y$ with joint probability distribution $p(X, Y)$, we define the smooth collision mutual information as
$$\mathsf{I}_c^\varepsilon(X : Y \mid Z) = \min_{X',Y',Z' \sim X,Y,Z} \mathsf{I}_c(X' : Y' \mid Z')$$
where the minimization is over all $(X', Y', Z')$ such that $\mathrm{SD}((X,Y,Z),(X',Y',Z')) \leq \varepsilon$.

**Lemma 3.1** (Standard Collision Information Facts): We have the following properties of collision information, similar to Shannon information:
1. **Additivity of independent variables:** $\mathsf{H}_c(X, Y \mid Z) = \mathsf{H}_c(X \mid Z) + \mathsf{H}_c(Y \mid Z)$ if $X$ and $Y$ are independent given $Z$
2. **Chain rule of mutual information:** $\mathsf{I}_c(X : YZ) = \mathsf{I}_c(X : Z) + \mathsf{I}_c(X : Y \mid Z)$
3. **Additivity of independent mutual information:** $\mathsf{I}_c(X, Y; Z, W \mid A) = \mathsf{I}_c(X; Z \mid A) + \mathsf{I}_c(Y; W \mid A)$ if $X$ and $Y$ are independent given $Z$ and $W$ when conditioned on $A$.
4. **Maximum of Collision Entropy:** $\mathsf{H}_c(X \mid Y) \leq \log_2|X|$ for all random variables $X$.

We will then use the above lemmas and smoothness to get the following:

**Lemma 3.2** (Upper-Bound on Collision Mutual Information): We have that $\mathsf{I}_c^\gamma(X : Y \mid Z) \leq \min(\log_2|X|, 2\log_2|Y|)$ and $\mathsf{I}_c(X : Y \mid Z) \leq \log_2|X|$ where $\gamma = \frac{1}{2\,|Y|}$.

**Lemma 3.3** (Collision Mutual Information is Nonnegative):

$$\mathsf{I}_c(X:Y) \geq 0$$
for all probability distributions $p$ over $X, Y$.

We also have the following useful convexity property as with Shannon information.

**Lemma 3.4** (Convexity of Collision Mutual Information): Let $p(X \mid Y) = \alpha q(X \mid Y) + (1 - \alpha)r(X \mid Y)$ for $\alpha \in [0,1]$ and $p(X) = q(X) = r(X)$. Then
$$\mathsf{I}_{c_p}(X:Y) \leq \alpha \mathsf{I}_{c_q}(X:Y) + (1-\alpha)\mathsf{I}_{c_r}(X:Y)$$
where $\mathsf{I}_{c_p}, \mathsf{I}_{c_q}, \mathsf{I}_{c_r}$ denotes the collision mutual information for distributions $p, q, r$ respectively.

### $\mathsf{I}_c$ is Cryptographically Relevant

We now provide our main technical reason for defining and using (smooth) collision mutual information. The below can be seen as a very simple application of Markov's inequality as in [37], which considers the left-over hash lemma in the context of collision entropy.

**Lemma 3.5** (Collision Mutual Information is Cryptographically Relevant): Let $X, Y, Z$ be random variables such that
$$\mathsf{I}_c(X:Z) \leq s - B$$
alongside the following: *if* $H_\infty(X \mid Z) \geq \mathsf{H}_c(X) - s$, *then*,
$$\mathsf{I}_c(X:Y \mid Z) \leq C.$$
Then,
$$\mathsf{I}_c^{2^{-B}}(X:Y \mid Z) \leq C$$

*Proof*: Consider that $\mathsf{I}_c(X:Z) \geq s - B$ implies that
$$\mathsf{H}_c(X \mid Z) = \mathsf{H}_c(X) - \mathsf{I}_c(X:Z) \geq \mathsf{H}_c(X) - s + B.$$
and so,
$$\mathbb{E}[P(X \mid Z)] \leq 2^{-\mathsf{H}_c(X)+s} \cdot 2^{-B}.$$
Then, by Markov's inequality, we have that
$$\Pr_{x,z}\left[P(X \mid Z) \geq 2^{-\mathsf{H}_c(X)+s}\right] \leq \frac{1}{2^B} = 2^{-B}$$
and
$$\Pr_{x,z}\left[P(X \mid Z) \leq 2^{-\mathsf{H}_c(X)+s}\right] \geq 1 - 2^{-B}.$$
We can then define a distribution $P'$ over $X, Y, Z$ such that $\Pr\left[P'(X \mid Z) \geq 2^{-\mathsf{H}_c(X)+s}\right] = 0$. Then, we have that
$$\mathrm{SD}(P, P') = \sup_A |P(A) - P'(A)| \leq 2^{-B}.$$
So,
$$\Rightarrow \mathsf{I}_c^{2^{-B}}(X:Y \mid Z) \leq -C$$
$$\Rightarrow \mathsf{I}_c^{2^{-B}}(X:Y \mid Z) \leq 2^{-B}\log_2|Y| + (1 - 2^{-B})\log_2 C$$

∎

# 4 Facts About $2 \mapsto 1$ Quantum Random Access Code

In this section, we will define the greater and total bit mutual information for a $2 \mapsto 1$ QRAC scheme and show how to calculate them. Specifically, these quantities will be useful for bounding the information leakage of a QRAC scheme when the input bits are independently and uniformly sampled. Missing proofs and details are deferred to the appendix (see Appendix B).

*Definition 4.1* (Greater and Total Bit Mutual Information): For a $2 \mapsto 1$ QRAC ensemble which maps uniformly and independently sampled bits $b_0, b_1 \in \{0,1\}$ to state $\mathcal{E}(b_0, b_1)$, the greater and total bit mutual information are defined as

$$I_> := \sup_\mu \max(\mathsf{I}_c[b_0 : \mu(\mathcal{E}(b_0, b_1))], \mathsf{I}_c[b_1 : \mu(\mathcal{E}(b_0, b_1))])$$

and

$$I_{\text{tot}} := \sup_\mu (\mathsf{I}_c[b_0 : \mu(\mathcal{E}(b_0, b_1))] + \mathsf{I}_c[b_1 : \mu(\mathcal{E}(b_0, b_1))]).$$

respectively. Moreover, we define a variant of greater bit mutual information, conditional bit mutual information, as follows

$$\tilde{I}_> := \sup_\mu \max\left(\mathsf{I}_c[b_0 : \mu(\mathcal{E}(b_0, b_1)) \mid b_1], \mathsf{I}_c[b_1 : \mu(\mathcal{E}(b_0, b_1)) \mid b_0]\right).$$

We also abuse notation to refer to $I_>$, $I_{\text{tot}}$, $\tilde{I}_>$ to be the respective constant of the greater, total, and conditional bit mutual informtation for the optimal $2 \mapsto 1$ QRAC scheme as defined in Definition 2.2.

We will also need the following lemma:

**Lemma 4.1** (Independence implies subadditivity): Given $\mathcal{E}(b_0^1, b_1^1), ..., \mathcal{E}(b_0^m, b_1^m)$, where $b_0^i, b_1^i$ for $i \in [m]$ are independently and uniformly sampled, then

$$\sup_\mu \max_{y \in \{0,1\}^m} \mathsf{I}_c\left(b_{y_0}^1 b_{y_1}^2 ... b_{y_m}^m : \mu\left(\bigotimes_{i \in [m]} \mathcal{E}(b_0^i, b_1^i)\right)\right) \leq \sum_{i \in [m]} I_> = m \cdot I_>$$

and

$$\sup_\mu \mathsf{I}_c\left(b_0^1 b_1^1 ... b_0^m b_1^m : \mu\left(\bigotimes_{i \in [m]} \mathcal{E}(b_0^i, b_1^i)\right)\right) \leq \sum_{i \in [m]} I_{\text{tot}} = m \cdot I_{\text{tot}}.$$

We also have a similar lemma for the subadditivity of conditional bit mutual information.

**Lemma 4.2**: For both $\alpha \in \{0,1\}$, let $b_\alpha^1, ..., b_\alpha^m$ be independently and uniformly sampled bits and $b_{1-\alpha}^1, ..., b_{1-\alpha}^m$ be arbitrary bits.

$$\sup_\mu \mathsf{I}_c\left(b_\alpha : \mu\left(\bigotimes_i \mathcal{E}(b_0^i, b_1^i) \,\middle|\, b_{1-\alpha}^1 ... b_{1-\alpha}^m\right)\right) \leq \sum_{i \in [m]} \tilde{I}_> = m \cdot \tilde{I}_>.$$

Finally, we get the following useful corollary from the prior two lemmas. Importantly, this corollary will be useful for bounding the information leakage of a QRAC scheme when the input bits are independently and uniformly sampled.

**Corollary 4.1** (Upper bound on greater string and total information): We can specialize the prior two lemmas to get, for independently and uniformly sampled bits $b_0^1, b_1^1, ..., b_0^m, b_1^m$,

$$\sup_\mu \max_{s \in \{0,1\}} \left[ \mathsf{I}_c \left( b_s^1 b_s^2 ... b_s^m : \mu \left( \bigotimes_{i \in [k+1]} \mathcal{E}(b_0^i, b_1^i) \right) \right) \right] \leq m \cdot I_>,$$

$$\sup_\mu \mathsf{I}_c \left( b_0, b_1 : \mu \left( \bigotimes_{i \in [k+1]} \mathcal{E}(b_0^i, b_1^i) \right) \right) \leq m \cdot I_{\text{tot}},$$

and, for independently and uniformly sampled bits $b_\alpha^1, ..., b_\alpha^m$ and arbitrary bits $b_{1-\alpha}^1, ..., b_{1-\alpha}^m$,

$$\sup_\mu \mathsf{I}_c \left( b_\alpha : \mu \left( \bigotimes_{i \in [k+1]} \mathcal{E}(b_0^i, b_1^i) \right) \bigg| b_{1-\alpha} \right) \leq m \cdot \tilde{I}_>.$$

## 4.1 Calculating $I_>$, $I_{\text{tot}}$, and $\tilde{I}_>$

We defer most of the details of the calculations to Appendix B.1. At a high level, we will find an upper-bound on $I_>, I_{\text{tot}}, \tilde{I}_>$ by maximizing over all possible **extremal** POVMs[6]. We then make use of Lemma 3.4, the convexity of collision mutual information, to get a bound on all possible POVMs.

To calculate $I_>, I_{\text{tot}}, \tilde{I}_>$ when restricting to extremal POVMs, we use an $\epsilon$-net approach where we discretize the set of extremal POVMs and brute force search to find a maximum value amongs the discretized set. We then make use of simple analytical techniques to get an upper-bound for all extremal POVMs given an upper-bound on the discretized set of POVMs.

Running the code[7], we get the following bounds:

**Fact 4.1** (Bounds on bit accessible mutual information): Given the $2 \mapsto 1$ QRAC scheme of Definition 2.2, we get that

$$I_> \leq 0.59$$
$$\tilde{I}_> \leq 0.59$$
$$I_{\text{tot}} \leq 0.65.$$

# 5 $\text{GQNC}_d^D$ and $C^{\text{1-GQNC}_d^D}$ Adversaries

In this section, we will formally define the complexity class $\text{GQNC}_d^D$ and $C^{\text{1-GQNC}_d^D}$. Here, $d$ refers to the maximum depth of the quantum circuit and $D$ refers to the *geometric* dimension of the circuit where we define a circuit's geometric dimension as follows. Though many papers have considered geometrically constrained circuits [38–41], we will define it here for clarity.

*Definition 5.1* (Geometric Dimension of a Circuit): Given a quantum circuit $C$ with $n$ qubits, we define the geometric dimension of the circuit as the smallest dimension $D$ such that there exists an equivalent circuit with gates acting on neighbors on qubits fixed within a $D$-dimensional grid, where every input qubit is a vertex in the grid.

---

[6]Recall that an extremal POVM is a POVM which cannot be written as a convex combination of other POVMs. For a $d$ dimensional subspace, an extremal POVM has at most $d^2$ elements.

[7]The code can be found on this Github repository.

As an example, we recall the illustration in Section 1 of a 2D geometric circuit in Figure 2.

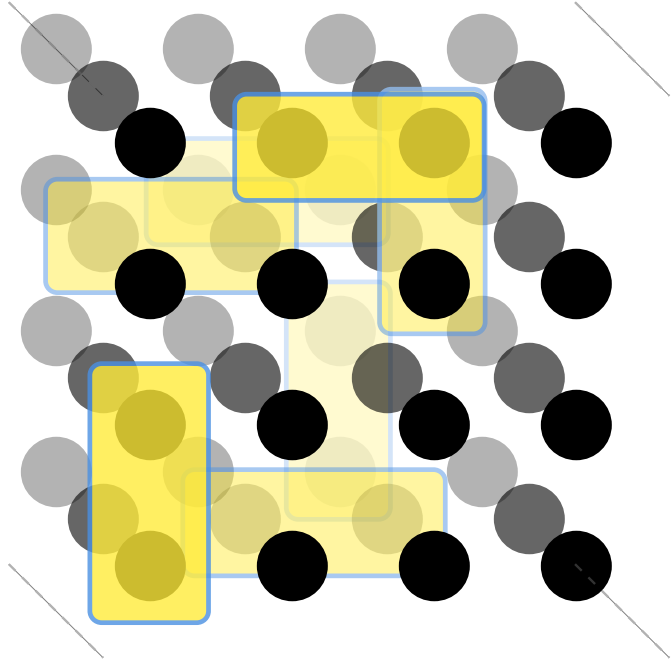

Figure 2: An example of a GQNC$_3^2$ circuit where we have a depth 3 circuit on a 2D grid of qubits. Each "layer" of the image represents a different time step in the circuit and each black dot represents a qubit within each layer. The yellow boxes represent the gates acting on a pair of qubits.

Then, we define the complexity class GQNC$_d^D$ as follows from the definition of QNC.

*Definition 5.2* (GQNC$_d^D$ circuit with $\ell$ local gates): A GQNC$_d^D$ circuit is a quantum circuit with $\ell$-local gates (where each gate acts on at most $\ell$ neighobring qubits) qubits and the circuit has depth at most $d$ and geometric dimension at most $D$. All measurements must be made at the end of the quantum circuit.

Finally, we define our adversary class as $C^{1\text{-GQNC}_d^D}$ as follows:

*Definition 5.3* ($C^{1\text{-GQNC}_d^D}$ Adversaries): A $C^{1\text{-GQNC}_d^D}$ adversary is an unbounded classical circuit which can make use of at most **one** GQNC$_d^D$ circuit of its choosing as a subroutine.

Note that, although a $C^{1\text{-GQNC}_d^D}$ adversary can only *perform* a single quantum circuit, it can simulate all possible quantum circuits and use the results in its computation. Thus, we argue that $C^{1\text{-GQNC}_d^D}$ captures the class of adversaries which can perform limited computation on *specific input/ auxiliary* quantum states while still being able to perform arbitrary (and even error-corrected) quantum computation on non-auxiliary states .

In this work, we take advantage of the constant depth and geometric locality of GQNC$_d^D$ by partitioning the input qubits into independent light cones which can be set to an arbitrary size.

Specifically, we will define a partition of the grid into cubes and then further partition within the cubes. We provide an example of such a partition in Figure 3.

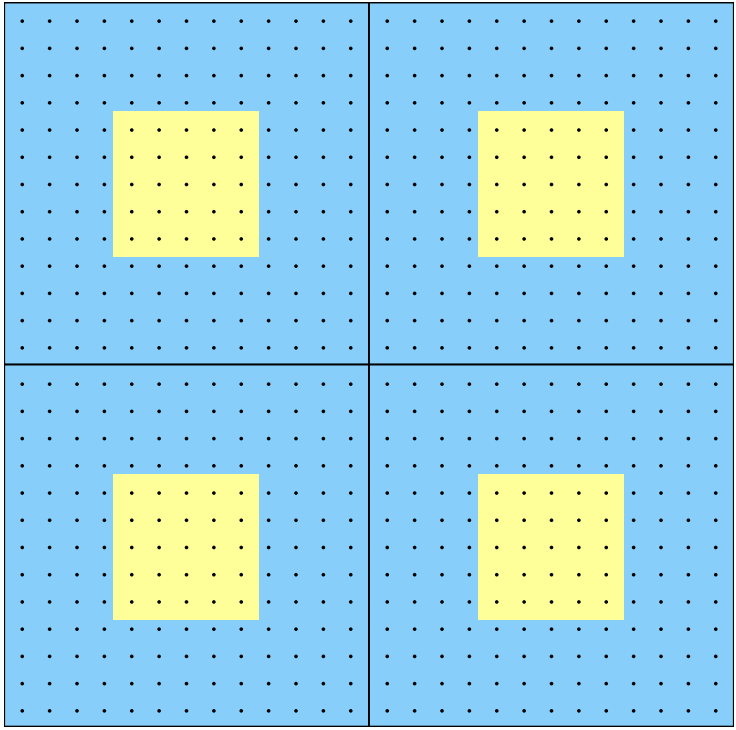

Figure 3: An example of a $2D$ grid of qubits partitioned into inner cubes and an outer shells for $\ell = 2, d = 2$, and geometric dimension of 2 ($D = 2$). Each larger square, outlined in black, represents a cube $\widetilde{\mathtt{cu}}_j$ with radius $r + \ell^d = 7$. Each blue shaded region represents the "shell," $\widetilde{\mathtt{sh}}_j$, of the cube $\widetilde{\mathtt{cu}}_j$ and each yellow region represents the "inner cube", $\mathtt{cu}_j$, of radius $r = 3$.

*Definition 5.4* (Hypercube Partition with radius $r$): Without loss of generality, assume that the number of input qubits, $n$, is a multiple of $(2r + 2\ell^d)^D$. For $q = n/(2r + 2\ell^d)^D$, partition the gird of qubits into "outer" hypercubes, $\widetilde{\mathtt{cu}}_1, ..., \widetilde{\mathtt{cu}}_q$ with radius $r + \ell^d$ and centers $c_1, ..., c_q$. Then, the inner cube $\mathtt{cu}_j$, for $j \in [q]$, is the set of qubits in the hybercube of radius $r$ and center $c_j$.

We then have the following lemma.

**Lemma 5.1** (Independence of Partitioning into Hypercubes): For all $j \in [q]$, any qubit outisde of $\widetilde{\mathtt{cu}}_j$ is not within the reverse light cone of depth $d$ originating from a qubit within the inner cube $\mathtt{cu}_j$.

*Proof*: The proof follows from a simple geometric observation. The qubits on the surface of $\mathtt{cu}_j$ are at most distance $r$ away from the center as the closest to all other qubits which are not inside $\mathtt{cu}_j$. Then, given that every qubit on the surface area of the cube has a light cone of diamater $\ell^d$, the qubits outside of $\widetilde{\mathtt{cu}}_j$ are not within the reverse light cone of depth $d$ originating from a qubit within the inner cube $\mathtt{cu}_j$. ∎

*Definition 5.5* (Further Partition): Let the "shell," $\widetilde{\mathtt{sh}}_j$ of a cube $\widetilde{\mathtt{cu}}_j$ equal $\widetilde{\mathtt{cu}}_j \setminus \mathtt{cu}_j$. Then let $\mathtt{CU} = \cup_{j=1}^q \mathtt{cu}_j$ and $\overline{\mathtt{CU}} = \cup_{j=1}^q \widetilde{\mathtt{sh}}_j = [n] \setminus \mathtt{CU}$.

**Lemma 5.2** (Independent Light Cones): Let input state, $\rho = \bigotimes_i \rho_i$, encode classical information $x$ for $\rho_i \in \mathbb{C}^2$. Then, for any measurement $\mu$,

$$\mathsf{I}_c^\epsilon(x:\mu(\rho)) \leq \mathsf{I}_c\left(x:\mu'(\rho^{\mathtt{CU}})\,\big|\,"\rho^{\overline{\mathtt{CU}}}"\right) + 2\cdot|"\rho^{\overline{\mathtt{CU}}}"|$$

where $\epsilon = 1/2^{|\overline{\mathtt{CU}}|}$ and $\mu'(x^{\mathtt{CU}})$ is the quantum circuit and measurement containing qubits in $\mathtt{CU}$ and $"\rho^{\overline{\mathtt{CU}}}"$ is the complete description of the state of qubits in $\overline{\mathtt{CU}}$.

*Proof*: First, by the chain rule of (collision) mutual information, we have that
$$\mathsf{I}_c(x:\mu(\rho)) \leq \mathsf{I}_c\left(x:\mu(\rho^{\mathtt{CU}})\,\big|\,"\rho^{\overline{\mathtt{CU}}}"\right) + \mathsf{I}_c\left(x:\rho^{\overline{\mathtt{CU}}}\right)$$
and then, as $\mathsf{I}_c^{\epsilon\left(x:\rho^{\overline{\mathtt{CU}}}\right)}$ is upper-bounded by the twice the log of the number of possible states of $\overline{\mathtt{CU}}$, as in Lemma 3.2, which is at most $2\cdot|"\rho^{\overline{\mathtt{CU}}}"|$, the length of the state description. ∎

We can now use the above language to regroup the adversary's measurements.

**Lemma 5.3** (Regrouping Qubit Measurement): Let $\mu_1...\mu_z$ be an ordered set of measurements made for $\mathrm{GQNC}_d^D$ circuit $C$ on qubits in $\mathtt{CU}$. Then,
$$\mathsf{I}_c\left(x:\mu_1,...,\mu_z\,\big|\,"\rho^{\overline{\mathtt{CU}}}"\right) = \mathsf{I}_c\left(x:\mu_1'...\mu_q'\,\big|\,"\rho^{\overline{\mathtt{CU}}}"\right)$$
where $\mu_j'$ represents the measurements made on qubits in $\mathtt{cu}_j$ for $j \in [q]$. We reorder the measurement such that for all $a < b$, $\mu_a'$ occurs prior to $\mu_b'$.

*Proof*: Because the circuit is non-adaptive, the order of measurements does not matter. So, we can regroup the measurements in any order and the measurement result will be the same. ∎

# 6 One-Time Memories for Random Messages

In this section, we will prove our main technical result. Mainly, we will show how to construct one-time memories for random strings, $r_0, r_1$, with bounds on the collision mutual information.

---

**Algorithm 1:** `Prep`: Creating the state with global codes, $\mathcal{C}_0 = (G_0, H_0), \mathcal{C}_1 = (G_1, H_1)$ where $\mathcal{C}_0, \mathcal{C}_1$ have rate $R \approx 0.4$ and probability $\epsilon_{\mathrm{corr}}$ of decoding failure for binary symmetric channel (BSC) with crossover probability $1 - \cos^2\left(\frac{\pi}{8}\right)$.

---

1 Sample random $r_0 \leftarrow 2^k$, $r_1 \leftarrow 2^k$ and set $c_0 = G_0 r_0$ and $c_1 = G_1 r_1$
2 Prepare $|\psi\rangle = \otimes_{i=1}^n \mathcal{E}(c_0^i, c_1^i)$
3 Return $|\psi\rangle, c_0, c_1$ and send $|\psi\rangle$ to the reciever

---

**Algorithm 2:** `Read`: Evaluating the OTM for string $c_\alpha$, $\alpha \in \{0,1\}$

---

1 On input $|\psi\rangle$, measure each qubit $i \in [n]$ using $\mu_\alpha$ to get output $c^i$
2 Let $c_\alpha' = (c^1,...,c^n)$
3 $c_\alpha \leftarrow \mathsf{Decode}(c_\alpha')$
4 Output $c_\alpha$

---

**Parameters**

Before diving into the protocol and proof, we need to outline the parameters used in the protocol and proof for easy reference:

- $\epsilon'$: an exponentially small constant in the security parameter which will be used to bound statistical distance.
- $\epsilon''$: an exponentially small constant in the security parameter which will be used to go from "average-case" to "worst-case" security.
- $n$: the number of qubits in the protocol as well as the length of codewords in the linear code
- $R$ the rate of the linear codes to be used where $\frac{7I_> + I_\leq}{8} + \epsilon'' < R \leq I_>$ as defined in Definition 4.1. We assume that $R$ is chosen such that $k = Rn$ is an integer.
- $\lambda$: a security parameter related to the length of the secret encoded inthe one-time memory as well as a bound on statistical distance. $\lambda$ can grow with $n$.
- $\ell, d, D$: the parameters of the $\text{GQNC}_d^D$ adversary as defined in Section 5.
- $r$: the radius of the inner cube in the hypercube partition as defined in Definition 5.4.
- $\mathtt{CU}, \overline{\mathtt{CU}}$: the inner cube and outer shell of the hypercube partition as defined in Definition 5.5.

We then have the following restriction on the above parameters:

$$2\,|\mathtt{cu}| + 2\log\left(\frac{1}{\epsilon'}\right) + \log\left(\frac{1}{\epsilon''}\right) + 4\,|\overline{\mathtt{CU}}| + (2r)^D \leq \frac{n}{100} \tag{1}$$

and

$$|\overline{\mathtt{CU}}| \geq \log\left(\frac{1}{\epsilon'}\right) \tag{2}$$

We prove the feasability of these bounds in Appendix C.

**Correctness**

> *Theorem 6.1* (Correctness): Following the measuring and decoding protocol outlined in Algorithm 1 and Algorithm 2, the output is $c_\alpha$ for chosen $\alpha \in \{0,1\}$ with probability $1 - \epsilon_{\text{corr}}$.

*Proof*: The proof follows simply from the fact that the $2 \mapsto 1$ QRAC measures the encoded bit with a probability of $\cos^2\left(\frac{\pi}{8}\right)$. So, we can think of the QRAC as a classical channel with a probability of error of $1 - \cos^2\left(\frac{\pi}{8}\right)$. Then, we can use the fact that the random linear codes $\mathcal{C}_0, \mathcal{C}_1$ have a decoding failure rate of $\epsilon_{\text{corr}}$ for binary symmetric channel $\text{BSC}\left(1 - \cos^2\left(\frac{\pi}{8}\right)\right)$. ∎

## 6.1 Average Case Soundness

*Definition 6.1* (Lesser Codeword): We say that a codeword $c_\alpha$ is the **lesser codeword** if
$$\mathsf{I}_c(c_\alpha : \mu(|\psi\rangle)) < \mathsf{I}_c(c_{1-\alpha} : \mu(|\psi\rangle))$$
for the adversary's measurement $\mu$.

Next, we state the main theorem of this section.

> *Theorem 6.2* (Average Case Soundness): Assuming that $c_1$ is the lesser codeword and that the adversary is in $C^{\text{1-GQNC}_d^D}$, then for uniformly random codeword $c_1$:
> $$\mathsf{H}_c^\epsilon(c_1 \mid \mu_{\mathcal{A}}(\mathcal{E}(c_0, c_1)), \mathcal{C}_0, \mathcal{C}_1) \geq \frac{\lambda}{2}$$

for some exponentially small $\epsilon$.

**Proof**

We will now prove the intermediary lemmas required to prove Theorem 6.2. Before we proceed, assume that $c_1$ is the lesser codeword. Recall the regrouping of measurements for a $\text{GQNC}_d^D$ circuit in Lemma 5.3: let $\mu'_1, ..., \mu'_q$ be the measurements on the qubits in $\texttt{cu}_1, ..., \texttt{cu}_q$ respectively. Then, using the regrouped measurements, we now bound how much information an adversary can learn about both codewords from measurement $\mu'_j$.

**Lemma 6.1** (Progress Measure for Mutual Information): Let $\mathcal{Y}_j = \cup_{i=1}^j \texttt{cu}_i$ and $\overline{\mathcal{Y}}_j = \texttt{CU} \setminus \mathcal{Y}_j$. Also, let $\mu\left(\mathcal{E}\left(c_0^{\mathcal{Y}_i}, c_1^{\mathcal{Y}_i}\right)\right) = \mu'_1 \cup \mu'_2 ... \cup \mu'_i$ be the adversary's measurements on the qubits in $\mathcal{Y}_j$ but regrouped as in Lemma 5.3. Then, if for both $\alpha \in \{0,1\}$,

$$\mathsf{I}_c^\epsilon\left(c_\alpha : \mu\left(\mathcal{E}\left(c_0^{\mathcal{Y}_{j-1}}, c_1^{\mathcal{Y}_{j-1}}\right)\right), c_0^{\overline{\mathcal{Y}}}, c_1^{\overline{\mathcal{Y}}}\right) \leq k - |\texttt{cu}_j| - 2\log\left(\frac{1}{\epsilon'}\right) - \log\left(\frac{1}{\epsilon''}\right), \tag{3}$$

we can bound the conditional mutual information after the $j$-th measurement as

$$\max_{\alpha \in \{0,1\}} \mathsf{I}_c^{\epsilon+\epsilon'+\epsilon''}\left(c_\alpha : \mu\left(\mathcal{E}\left(c_0^{\mathcal{Y}_j}, c_1^{\mathcal{Y}_j}\right)\right) \,\middle|\, \mu\left(\mathcal{E}\left(c_0^{\mathcal{Y}_j}, c_1^{\mathcal{Y}_j}\right)\right), c_0^{\overline{\mathcal{Y}}}, c_1^{\overline{\mathcal{Y}}}\right) \leq I_> \cdot |\texttt{cu}_j|$$

and

$$\mathsf{I}_c^{\epsilon+\epsilon'+\epsilon''}\left(c_0, c_1 : \mu\left(\mathcal{E}\left(c_0^{\mathcal{Y}_j}, c_1^{\mathcal{Y}_j}\right)\right) \,\middle|\, \mu\left(\mathcal{E}\left(c_0^{\mathcal{Y}_{j-1}}, c_1^{\mathcal{Y}_{j-1}}\right)\right), c_0^{\overline{\mathcal{Y}}}, c_1^{\overline{\mathcal{Y}}}\right) \leq I_{\text{tot}} \cdot |\texttt{cu}_j|.$$

*Proof*: First note that

$$\mathsf{H}_c^\epsilon\left(c_\alpha \,\middle|\, \mu\left(\mathcal{E}\left(c_0^{\mathcal{Y}_{j-1}}, c_1^{\mathcal{Y}_{j-1}}\right)\right), c_0^{\overline{\texttt{CU}}}, c_1^{\overline{\texttt{CU}}}\right) \geq \mathsf{H}_c^\epsilon(c_\alpha) - \mathsf{I}_c^\epsilon\left(c_\alpha : \mu\left(\mathcal{E}\left(c_0^{\mathcal{Y}_{j-1}}, c_1^{\mathcal{Y}_{j-1}}\right)\right), c_0^{\overline{\texttt{CU}}}, c_1^{\overline{\texttt{CU}}}\right)$$

$$\geq k - \left(k - |\texttt{cu}_j| - 2\log\left(\frac{1}{\epsilon'}\right) - \log\left(\frac{1}{\epsilon''}\right)\right)$$

$$= |\texttt{cu}_j| + 2\log\left(\frac{1}{\epsilon'}\right) + \log\left(\frac{1}{\epsilon''}\right)$$

by definition of collision mutual information. If

$$H_\infty^\epsilon\left(c_\alpha \,\middle|\, \mu\left(\mathcal{E}\left(c_0^{\mathcal{Y}_{j-1}}, c_1^{\mathcal{Y}_{j-1}}\right)\right), c_0^{\overline{\texttt{CU}}}, c_1^{\overline{\texttt{CU}}}\right) \geq |\texttt{cu}_j| + 2\log\left(\frac{1}{\epsilon'}\right) \tag{4}$$

and $h$ is a 2-universal hash function, $h(c_\alpha)$ is $\epsilon'$ close to uniform for by the left over hash lemma. We can see that $G_\alpha$, by definition is a random matrix and thus $G_\alpha[\texttt{cu}_j]$, defined as the rows of $G_\alpha$ indexed by $\texttt{cu}_j$, is a random matrix as well in $\{0,1\}^{n \times |\texttt{cu}_j|}$. We can then see that $G_\alpha[\texttt{cu}_j]$ is a 2-universal hash function and so,

$$\mathrm{SD}(c_\alpha \cdot G_\alpha[\texttt{cu}_j], \mathrm{Uniform}) \leq \epsilon'$$

by the left over hash lemma when given $\mu\left(\mathcal{E}\left(c_0^{\mathcal{Y}_{j-1}}, c_1^{\mathcal{Y}_{j-1}}\right)\right), c_0^{\overline{\texttt{CU}}}, c_1^{\overline{\texttt{CU}}}$. Moreover, we can note that $I_>, I_{\text{tot}}$ bounds greater and total mutual information for a measurement on uniformly random QRAC states. Recall Corollary 4.1 gives us bounds on the mutual information for the greater and total bit strings in uniformly random QRAC states. So, under the condition that

$$H_\infty^\epsilon\left(c_\alpha \,\middle|\, \mu\left(\mathcal{E}\left(c_0^{\mathcal{Y}_{j-1}}, c_1^{\mathcal{Y}_{j-1}}\right)\right), c_0^{\overline{\texttt{CU}}}, c_1^{\overline{\texttt{CU}}}\right) \geq |\texttt{cu}_j| + 2\log\left(\frac{1}{\epsilon'}\right),$$

we have that $c_\alpha \cdot G_\alpha[\texttt{cu}_j]$ is close to IID and thus

$$\mathsf{I}_c^{\epsilon+\epsilon'}\Big(c_\alpha : \mu\big(\mathcal{E}(c_0^{\mathtt{cu}_j}, c_1^{\mathtt{cu}_j})\big) \,\Big|\, \mu\big(\mathcal{E}(c_0^{\mathcal{Y}_j}, c_1^{\mathcal{Y}_j})\big), c_0^{\overline{\mathtt{CU}}}, c_1^{\overline{\mathtt{CU}}}\Big) \leq I_> |\mathtt{cu}_j|$$

and

$$\mathsf{I}_c^{\epsilon+\epsilon'}\Big(c_0, c_1 : \mu\big(\mathcal{E}(c_0^{\mathtt{cu}_j}, c_1^{\mathtt{cu}_j})\big) \,\Big|\, \mu\big(\mathcal{E}(c_0^{\mathcal{Y}_j}, c_1^{\mathcal{Y}_j})\big), c_0^{\overline{\mathtt{CU}}}, c_1^{\overline{\mathtt{CU}}}\Big) \leq I_{\text{tot}} |\mathtt{cu}_j|$$

by Corollary 4.1. Thus, if we assume that eq. (4) holds for both $\alpha \in \{0,1\}$, we have that the qubits in $\mathtt{cu}_j$ have high conditional collision entropy. Finally, we make use of the cryptographic properties of collision mutual information (Lemma 3.5), to get that

$$\mathsf{I}_c^{\epsilon+\epsilon'+\epsilon''}\Big(c_\alpha : \mu\big(\mathcal{E}(c_0^{\mathtt{cu}_j}, c_1^{\mathtt{cu}_j})\big)\Big) \,\Big|\, \mu\big(\mathcal{E}(c_0^{\mathcal{Y}_j}, c_1^{\mathcal{Y}_j})\big), c_0^{\overline{\mathtt{CU}}}, c_1^{\overline{\mathtt{CU}}}\Big) \leq I_> |\mathtt{cu}_j|$$

and

$$\mathsf{I}_c^{\epsilon+\epsilon'+\epsilon''}\Big(c_0 c_1 : \mu\big(\mathcal{E}(c_0^{\mathtt{cu}_j}, c_1^{\mathtt{cu}_j})\big)\Big) \,\Big|\, \mu\big(\mathcal{E}(c_0^{\mathcal{Y}_j}, c_1^{\mathcal{Y}_j})\big), c_0^{\overline{\mathtt{CU}}}, c_1^{\overline{\mathtt{CU}}}\Big) \leq I_{\text{tot}} |\mathtt{cu}_j|$$

as desired. ∎

**Corollary 6.1** (Expanding Recursion for Greater Codeword): If we exapnd the recursive inequality in Lemma 6.1 and $\alpha = 0$, we get that

$$\mathsf{H}_c^\epsilon\Big(c_0 \,\Big|\, \mu\big(\mathcal{E}(c_0^{\mathcal{Y}_j}, c_1^{\mathcal{Y}_j})\big), c_0^{\overline{\mathtt{CU}}}, c_1^{\overline{\mathtt{CU}}}\Big) \geq k - I_> |\mathcal{Y}_j| - 4|\overline{\mathtt{CU}}|.$$

for exponentially small $\epsilon$.

*Proof*: The corollary follows directly from Lemma 6.1, except that we now consider the auxiliary information from $c_0^{\overline{\mathtt{CU}}}, c_1^{\overline{\mathtt{CU}}}$. Note that its classical description requires $2|\overline{\mathtt{CU}}|$ bits and thus by Lemma 5.2, we have that conditioning on $c_0^{\overline{\mathtt{CU}}}, c_1^{\overline{\mathtt{CU}}}$ leaks at most $4|\overline{\mathtt{CU}}|$ bits of information for distribution $1/2^{|\overline{\mathtt{CU}}|}$ close to the original distribution. Given that $|\overline{\mathtt{CU}}| \geq \log(\frac{1}{\epsilon'})$, we have that the bound is exponentially small as per eq. (2). ∎

To then get a bound on the information learned for the lesser codeword, a bit more work is required as we have to simultaneously use a few different bounds. We introduce the following lemma to bound the information learned about the lesser codeword after the adversary has learned most of the greater codeword.

**Lemma 6.2**: For some $j \in [q]$, if

$$\mathsf{I}_c^\epsilon\Big(c_0 : \mu\big(\mathcal{E}(c_0^{\mathcal{Y}_{j-1}}, c_1^{\mathcal{Y}_{j-1}})\big) \,\Big|\, c_0^{\overline{\mathtt{CU}}}, c_1^{\overline{\mathtt{CU}}}\Big) > k - |\mathtt{cu}_j| - 2\log\left(\frac{1}{\epsilon'}\right) - \log\left(\frac{1}{\epsilon''}\right),$$

and

$$\mathsf{I}_c^\epsilon\Big(c_1 : \mu\big(\mathcal{E}(c_0^{\mathcal{Y}_{j-1}}, c_1^{\mathcal{Y}_{j-1}})\big) \,\Big|\, c_0^{\overline{\mathtt{CU}}}, c_1^{\overline{\mathtt{CU}}}\Big) \leq k - |\mathtt{cu}_j| - 2\log\left(\frac{1}{\epsilon'}\right) - \log\left(\frac{1}{\epsilon''}\right), \tag{5}$$

then

$$\mathsf{I}_c^{\epsilon+\epsilon'+\epsilon''}\Big(c_1 : \mu\big(\mathcal{E}(c_0^{\mathtt{cu}_j}, c_1^{\mathtt{cu}_j})\big) \,\Big|\, \mu\big(\mathcal{E}(c_0^{\mathcal{Y}_{j-1}}, c_1^{\mathcal{Y}_{j-1}})\big), c_0^{\overline{\mathtt{CU}}}, c_1^{\overline{\mathtt{CU}}}\Big) \leq \tilde{I}_> \cdot |\mathtt{cu}_j|$$

*Proof*: The proof follows from the same reasoning as the proof of Lemma 6.1. The only difference is that the left-over hash lemma cannot be used on the greater code-word. Thus, we use the left-over hash lemma only on the lesser code-word (to get that it is close to uniform) and then assume that the adversary has learned the greater codeword completely. We then make use of our bound on conditional bit mutual information to get that the adversary learns at most $\tilde{I}_>$ bits of information about the lesser codeword per qubit in $\mathtt{cu}_j$. ∎

We are now ready to combine the above lemmas to get a bound on the (collision) mutual information learned about the lesser codeword.

Let $m \in [q]$ be such that the value of qubits insie the inner cube $\mathtt{cu}_i$, for $i \in [m]$, are indistinguishable from uniformly random and independent until at least measurement $\mu'_1, ..., \mu'_m$ are made: i.e. $m$ is the largest value such that eq. (3) holds for both $\alpha = 0, 1$.

**Lemma 6.3** (Upperbounds on lesser codeword mutual information): Assume that the measurement $\mu$ learns most of $c_0$:
$$\mathsf{I}^\nu_c(c_0 : \mu(\mathcal{E}(c_0^{\mathtt{CU}}, c_1^{\mathtt{CU}}))) > k - 4|\overline{\mathtt{CU}}| - \lambda \tag{6}$$
for some exponentially small $\nu$. Then, if we expand the recursive inequality in Lemma 6.1 and the above lemma, we get that
$$\mathsf{I}^\epsilon_c\left(c_1 : \mu(\mathcal{E}(c_0^{\mathtt{CU}}, c_1^{\mathtt{CU}})) \,\middle|\, c_0^{\overline{\mathtt{CU}}}, c_1^{\overline{\mathtt{CU}}}\right) \leq 0.31n$$
for exponentially small $\epsilon$.

*Proof*: First, note that we can write the information learned about $c_1$ in three steps:
1. The information learned about $c_1$ in the first $m$ measurements and from qubits in $\overline{\mathtt{CU}}$, denoted $M_1$
2. The information learned about $c_1$ in the next set of measurements where eq. (5) is satisfied, denoted $M_2$
3. The remaining measurements, denoted $M_3$

Also let $G$ be the information learned about $c_0$ in the first $m$ measurements and from the qubits in $\overline{\mathtt{CU}}$.

In the first $m$ set of measurements on $|\mathtt{cu}| \cdot m$ qubits, we can use Lemma 6.1 to get that
$$G + M_1 \leq |\mathtt{cu}| \cdot m \cdot I_{\text{tot}} + 4\,|\overline{\mathtt{CU}}|$$
where the value comes from the fact that the adversary leanrs at most $|\mathtt{cu}| \cdot m \cdot I_{\text{tot}}$ total bits of information. We also have the constraint, by definition of $m$, that
$$G \geq k - |\mathtt{cu}| - 2\log\left(\frac{1}{\epsilon'}\right) - \log\left(\frac{1}{\epsilon''}\right).$$
Then, for the next $M_2$ measurements, we can use Lemma 6.2 to get that
$$M_2 \leq |\mathtt{cu}| \cdot (q - m) \cdot \tilde{I}_>.$$
as we have at most $q - m$ remaining cubes to measure and Lemma 6.2 gives us that the adversary learns at most $\tilde{I}_>$ bits of information about $c_1$ per qubit in $\mathtt{cu}_j$. Next, note that we can re-write a bound on $M_1 + M_2$ as
$$M_1 + G + M_2 \leq |\mathtt{cu}| \cdot m \cdot I_{\text{tot}} + 4\,|\overline{\mathtt{CU}}| + |\mathtt{cu}| \cdot (q - m) \cdot \tilde{I}_>$$
$$\Rightarrow M_1 + M_2 \leq |\mathtt{cu}| \cdot q \cdot I_{\text{tot}} + 4\,|\overline{\mathtt{CU}}| + |\mathtt{cu}| \cdot (q - m) \cdot \tilde{I}_> \tag{7}$$
$$- \left(k - |\mathtt{cu}| - 2\log\left(\frac{1}{\epsilon'}\right) - \log\left(\frac{1}{\epsilon''}\right)\right).$$
Recall the constraints in eq. (1),
$$4|\overline{\mathtt{CU}}| + 2|\mathtt{cu}| + 2\log\left(\frac{1}{\epsilon'}\right) + \log\left(\frac{1}{\epsilon''}\right) < \frac{n}{100}$$
and so,

$$M_1 + M_2 \leq |\mathtt{cu}| \cdot m \cdot I_{\text{tot}} + |\mathtt{cu}| \cdot (q-m) \cdot \tilde{I}_> + \frac{n}{100} - k$$
$$\approx 0.7\,|\mathtt{cu}| \cdot m + 0.6\,|\mathtt{cu}| \cdot (q-m) \cdot \tilde{I}_> + \frac{n}{100} - 0.4n$$
$$\leq 0.3n + \frac{n}{100} \quad (\text{as } |\mathtt{cu}| \cdot m + |\mathtt{cu}| \cdot (q-m) \leq n)$$
$$= 0.31n$$

where we recall the bounds from Section 4.1 yielding $I_>, \tilde{I}_>, I_{\text{tot}} \leq 0.7$. Then note that $M_3 = 0$ as eq. (5) is always satisfied for all measurements made on $\mathtt{cu}_{m+1}, \mathtt{cu}_{m+2}, ..., \mathtt{cu}_q$. Finally, we get our desired result as
$$\mathsf{I}_c^\epsilon(c_1 : \mathcal{E}(c_0, c_1)) = M_1 + M_2 + M_3 \leq 0.31n$$
for exponentially small $\epsilon$. ∎

We are now ready to prove Theorem 6.2.

*Proof of Theorem 6.2*: We know that
$$\mathsf{I}_c^\epsilon\Big(c_1 : \mu(\mathcal{E}(c_0^{\mathtt{CU}}, c_1^{\mathtt{CU}})), c_0^{\overline{\mathtt{CU}}}, c_1^{\overline{\mathtt{CU}}}\Big) \leq 0.31n.$$

Then, $\mathsf{H}_c(c_1) = k \approx 0.4n$ and so, we have that, for large enough $n$
$$\mathsf{H}_c^\epsilon\Big(c_1 \mid \mu(\mathcal{E}(c_0^{\mathtt{CU}}, c_1^{\mathtt{CU}})), c_0^{\overline{\mathtt{CU}}} c_1^{\overline{\mathtt{CU}}}\Big) \geq \mathsf{H}_c^\epsilon(c_1 \mid \mu(\mathcal{E}(c_0, c_1)))$$
$$\geq k - 0.31n \geq 0.09n \geq \frac{\lambda}{2}$$

as desired. ∎

## 7 One-Time Memories for All Messages

Finally, we can use the one-time random memories outlined in Section 6 to construct a fully secure one-time memory protocol against $C^{\text{1-GQNC}_d^D}$. We refer to Prep and Read in Algorithm 1 and Algorithm 2 respecitevly as OTRM.Prep and OTRM.Read.

Then, we can construct a fully secure one-time memory protocol as follows:

---
**Algorithm 3:** $\mathtt{Prep}(m_0, m_1)$ for $m_0, m_1 \in \{0,1\}^{\frac{\lambda}{8}}$: Preparing the state with global codes, $\mathcal{C}_0 = (G_0, H_0), \mathcal{C}_1 = (G_1, H_1)$

---

1. Sample $c_0, c_1, |\psi\rangle \leftarrow \mathtt{OTRM.Prep}(\lambda)$
2. Sample public randomness $\omega_0, \omega_1 \in \{0,1\}^\ell$
3. Sample $\mathtt{Ext}_\alpha(x) = \mathtt{Ext}(\omega_\alpha, x)$ for $\mathtt{Ext}: \{0,1\}^\ell \times \{0,1\}^n \to \{0,1\}^{\frac{\lambda}{8}}$
4. Let $\mathtt{ct}_0 = m_0 \oplus \mathtt{Ext}_0(c_0)$ and $\mathtt{ct}_1 = m_1 \oplus \mathtt{Ext}_1(r_1)$
5. Output $\mathtt{ct}_0, \mathtt{ct}_1$ and $|\psi\rangle, \mathtt{Ext}_0, \mathtt{Ext}_1$

---
**Algorithm 4:** Read: $\mathtt{ct}_0, \mathtt{ct}_1, |\psi\rangle, \mathtt{Ext}_0, \mathtt{Ext}_1, \alpha$

---

1. Call $r_\alpha \leftarrow \mathtt{OTRM.Read}(|\psi\rangle, \alpha)$ and let $s_\alpha = \mathtt{Ext}_\alpha(r_\alpha)$
2. Output $m_\alpha = s_\alpha \oplus \mathtt{ct}_\alpha$

## 7.1 Correctness

The correctness of the protocol is a straightforward extension of the correctness of the one-time random memories in Section 6.

*Theorem 7.1* (Correctness of One-Time Memory): The protocol in Algorithm 3 and Algorithm 4 is correct as defined in Definition 2.9 with probability $1 - \epsilon_{\text{corr}}$.

*Proof*: The proof of correctness follows directly from the correctness of the one-time random memories in Section 6. ∎

## 7.2 Security

We finally show that the protocol is secure against $C^{\text{1-GQNC}_d^D}$ adversaries as defined in Definition 2.10.

*Theorem 7.2* (Soundness of One-Time Memory): The protocol in Algorithm 3 and Algorithm 4 is secure against $C^{\text{1-GQNC}_d^D}$ adversaries as defined in Definition 2.10.

*Proof*: Without loss of generality, assume that $\alpha = 0$. We can now build a simulator $\texttt{Sim}$ as follows:
- $\texttt{Sim}$ recieves $g^{m_0,m_1}(0) = m_0$
- Sample $c_0, c_1, |\psi\rangle \leftarrow \texttt{OTRM.Prep}(\lambda)$ and $\texttt{Ext}_0, \texttt{Ext}_1$
- Sample $\texttt{ct}_1' \leftarrow \{0,1\}^{\frac{\lambda}{8}}$
- Set $\texttt{ct}_0' = m_\alpha \oplus \sum_j \texttt{Ext}_\alpha^j(c_\alpha^j)$
- Call $\mathcal{A}(|\psi\rangle, \texttt{ct}_0, \texttt{ct}_1, \texttt{Ext}_0, \texttt{Ext}_1)$

We first need to show that $\texttt{ct}_1$ in the real protocol is indistinguishable from $\texttt{ct}_1'$ in the simulation. Note $\texttt{OTRM.Prep}$ in the simulator is called in the same way as in the real protocol and so by Theorem 6.2

$$\mathsf{H}_c^\epsilon(c_1 \mid |\psi\rangle_j) \geq \frac{\lambda}{2}$$

where $\epsilon$ is exponentially small in $\lambda$. Moreover, note that $\texttt{Ext}_0, \texttt{Ext}_1$ are independent of $c_0, c_1$ in both the real protocol and the simulated protocol. So,

$$\mathsf{H}_c^\epsilon(c_1 \mid |\psi\rangle, \texttt{ct}_0, \texttt{ct}_1, \texttt{Ext}_0, \texttt{Ext}_1) \geq \mathsf{H}_c^\epsilon(c_1 \mid |\psi\rangle, \texttt{ct}_0, \texttt{ct}_1)$$
$$\geq \mathsf{H}_c^\epsilon(c_1 \mid |\psi\rangle) - |\texttt{ct}_0| - |\texttt{ct}_1|$$
$$\geq \frac{\lambda}{2} - |\texttt{ct}_0| - |\texttt{ct}_1|.$$

Next, by definition of collision conditional mutual information, we have that

$$\Pr_{c_1, |\psi\rangle, \texttt{ct}_0, \texttt{ct}_1}\left[\Pr_c[c_1 = c \mid |\psi\rangle, \texttt{ct}_0, \texttt{Ext}_0, \texttt{Ext}_1] \leq 2^{-\frac{\lambda}{4}}\right] \geq 1 - \frac{\mathbb{E}[\Pr_c[c_1 = c \mid |\psi\rangle, \texttt{ct}_0, \texttt{Ext}_0, \texttt{Ext}_1]]}{2^{-\frac{\lambda}{4}}}$$
$$\geq 1 - \frac{2^{-\frac{\lambda}{2}}}{2^{-\frac{\lambda}{4}}}$$
$$\geq 1 - 2^{-\frac{\lambda}{4}}$$

where the first inequality follows from Markov's inequality. Thus, with probability at least $1 - 2^{-\frac{\lambda}{4}}$, $c_1$ has min-entropy at least $\lambda/4$ in the real protocol and so, with probability at least $1 - 2^{-\frac{\lambda}{4}}$, $\texttt{Ext}_1(c_1)$ is indistinguishable from random and thus indistinguishable from $\texttt{ct}'_1$. Given that the remainder of the real and simulator protocols are identical, we have that the real and simulated protocols are indistinguishable. ∎

# 8 Discussion and Outlook

In this paper, we begin the exploration of statistically sound one-time memories in the presence of hardware constraints. Specifically, we show that one-time memories are possible assuming that the adversary is constrained to a non-adaptive, constant depth, and geometrically-local quantum circuit. Our results have a few key limitations though, mainly that we do not run in polynomial time and that we assume that the adversary is non-adaptive and geometrically-local constrained.

So, there are a few immediate follow-up questions:
- Can we use polynomial-time algorithms to achieve the same result? Specifically, can we use poly-time decodable codes to achieve the same result using a similar proof technique?
- Can we remove either the requirement for a non-adaptive quantum circuit or the geometrically-local constraints on the adversary?

While we do not have immediate answers to these questions, we believe that they are interesting directions for future work.

Moreover, even if we can use polynomial time decodable codes, the constants in the protocol may be too large to be practical. We believe that more fine-grained techniques for analyzing the adversary's progress in the protocol may be necessary to achieve practical protocols.

## Acknowledgments

The author is grateful to the helpful discussions and feedback from Matthew Coudron, Yi-Kai Liu, Stefano Gogioso, Shi Jie Samuel Tan, Fabrizio Romano Genovese, and Gorjan Alagic. The author also acknowledges funding and support from NeverLocal Ltd, Neon Tetra LLC, and from the NSF Graduate Research Fellowship Program.

# A) Proofs for Collision Information Theory

In this appendix, we provide missing proofs for the main text on Collision Information Theory. First, we will prove Lemma 3.1, restated below.

**Restated Lemma** (Collision Information Facts): We have the following properties of collision information, similar to Shannon information:
1. **Additivity of independent variables:** $\mathsf{H}_c(X, Y \mid Z) = \mathsf{H}_c(X \mid Z) + \mathsf{H}_c(Y \mid Z)$ if $X$ and $Y$ are independent given $Z$
2. **Chain rule of mutual information:** $\mathsf{I}_c(X : YZ) = \mathsf{I}_c(X : Z) + \mathsf{I}_c(X : Y \mid Z)$
3. **Additivity of independent mutual information:** $\mathsf{I}_c(X, Y; Z, W) = \mathsf{I}_c(X; Z) + \mathsf{I}_c(Y; W)$ if $X$ and $Y$ are independent given $Z$ and $W$
4. **Maximum of Collision Entropy:** $\mathsf{H}_c(X \mid Y) \leq \log_2 |X|$ for all random variables $X$.

*Proof*: We will prove the following without the logarithm, translating $+$ to $\times$, for simplicity.

1. Follows from the fact that $\Pr[X, Y \mid Z] = \Pr[X \mid Z] \cdot \Pr[Y \mid Z]$:

$$\mathbb{E}_{\{x,y,z\}}[\Pr[X = x, Y = y \mid Z = z]] = \sum_{x,y,z} \Pr[X = x, Y = y, Z = z] \cdot \Pr[X = x, Y = y \mid Z = z]$$

$$= \sum_z \Pr[Z = z]\left(\sum_{x,y} \Pr[X = x Y = y \mid Z = z]^2\right)$$

$$= \sum_z \Pr[Z = z]\left(\sum_x \Pr[X = x \mid Z = z]^2\right) \cdot \left(\sum_y \Pr[X = y \mid Z = z]^2\right)$$

$$= \mathbb{E}_x[\Pr[X = x \mid Z = z]] \cdot \mathbb{E}_y[\Pr[Y = y \mid Z = z]].$$

2. Follows from definition:
$$\frac{\mathbb{E}_{x,z,y} \Pr[X \mid YZ]}{\mathbb{E}_x \Pr[X]} = \frac{\mathbb{E}_{x,z} \Pr[X \mid Z]}{\mathbb{E}_x \Pr[X]} \cdot \frac{\mathbb{E}_{y,z} \Pr[X \mid YZ]}{\mathbb{E}_y \Pr[X \mid Z]}.$$

3. We first have
$$P(XW \mid YZ) = \frac{P(XW, YZ)}{P(YZ)}$$
$$= \frac{P(X,Y) \cdot P(W,Z)}{P(Y) \cdot P(Z)}$$
$$= \left(\frac{P(X,Y)}{P(Y)}\right) \cdot \left(\frac{P(W,Z)}{P(Z)}\right)$$
$$= P(X \mid Y) \cdot P(W \mid Z).$$

So then,
$$\frac{\mathbb{E}_{x,y,w,z} \Pr[X, W \mid Y, Z]}{\mathbb{E}_{x,y} \Pr[X, W]} = \frac{\mathbb{E}_{x,y} \Pr[X \mid Y]}{\mathbb{E} \Pr[X]} \cdot \frac{\mathbb{E}_{x,y} \Pr[W \mid Z]}{\mathbb{E} \Pr[W]}.$$

4. First, by Lemma 3.3, we have that $\mathsf{I}_c(X : Y) \geq 0$ and so $\mathsf{H}_c(X \mid Y) \leq \mathsf{H}_c(X)$. So then, we need to upper-bound $\mathsf{H}_c(X)$ for probability distribution $P$, $\max_P \mathsf{H}_{c\,P}(X) = \min_P \mathbb{E}_P[\Pr[X = x]]$. Then, we will use Cauchy-Schwartz in the form $(\sum a_i b_i)^2 \leq (\sum a_i^2)(\sum b_i^2)$ and let $a_i = \Pr[X = x_i]$ and $b_i = 1$. Then,

$$\left(\sum_{x \in X} \Pr[X = x]\right)^2 \leq \sum_{i \in [\![X]\!]} 1^2 \cdot \left(\sum_{x \in X} \Pr[X = x]^2\right)$$

$$= |X| \sum_{x \in X} \Pr[X = x]^2$$

$$\Rightarrow 1 \leq |X| \sum_{x \in X} \Pr[X = x]^2$$

$$\Rightarrow \frac{1}{|X|} \leq \sum_{x \in X} \Pr[X = x]^2 = \mathbb{E}[\Pr[X]]$$

as desired.

∎

Now, we prove Lemma 3.2:

**Restated Lemma** (Upper-Bound on Collision Mutual Information): We have that $\mathsf{I}_c^{\frac{1}{2|Y|}}(X : Y \mid Z) \leq \min(\log_2 |X|, 2\log_2 |Y|)$.

*Proof*: First, note that
$$\mathsf{I}_c(X : Y \mid Z) = \mathsf{H}_c(X \mid Z) - \mathsf{H}_c(X \mid YZ) \leq \mathsf{H}_c(X \mid Z) \leq \log_2 |X|.$$

Then, for distribution $P$ over $X, Y$, let $P'$ be the closest distribution to $P$ such that $\min_y P'(Y = y) \geq \frac{1}{|Y|^2}$. Thus, we have that

$$\text{SD}(P(Y), P'(Y)) \leq \frac{1}{2} \sum_y |P(Y=y) - P'(Y=y)| \leq \frac{1}{2} \frac{|Y|}{|Y|^2} = \frac{1}{2|Y|}.$$

Then note that,
$$\sum_y P'(X = x \mid Y = y) P'(Y = y) = P'(X = x)$$

$$\Rightarrow \min_y \ |Y| \cdot P'(X = x \mid Y = y) P'(Y = y) \leq P'(X = x)$$

$$\Rightarrow P'(X = x \mid Y = y) \leq \frac{P'(X = x)}{|Y| \cdot \min P'(Y = y)}$$

$$\Rightarrow P'(X = x \mid Y = y) \leq |Y| \ P'(X = x).$$

So then, we have
$$\frac{\mathbb{E}[P'(X \mid Y) P'(X, Y))]}{\mathbb{E}[P'(X)]} = \frac{\sum_{x,y} P'(X = x, Y = y) P'(X = x \mid Y = y)}{\sum_x P(X = x)^2}$$

$$\leq \frac{\sum_{x,y} P'(Y = y \mid X = x) P'(X = x) P'(X = x \mid Y = y)}{\sum_x P(X = x)^2}$$

$$\leq |Y| \frac{\sum_{x,y} P'(Y = y \mid X = x) P'(X = x)^2}{\sum_x P(X = x)^2}$$

$$\leq |Y| \sum_{x,y} P'(Y = y \mid X = x) \frac{\sum_x P'(X = x)^2}{\sum_x P(X = x)^2}$$

$$\leq |Y| \sum_{x,y} P'(Y = y \mid X = x) = |Y| \sum_y 1 = |Y|^2.$$

And so, we have that $\mathsf{I}_c^{\frac{1}{2|Y|}}(X : Y \mid Z) \leq \min(\log_2 |X|, 2\log_2 |Y|)$. ∎

Next, we will prove Lemma 3.3, restated below.

**Restated Lemma** (Collision Mutual Magnitude is Nonnegative):
$$\mathsf{I}_c(X:Y) \geq 0$$
for all probability distributions $p$ over $X, Y$.

*Proof*:
$$\mathbb{E}_{x,y}P(X|Y) = \sum_y P(Y=y) \sum_x P(X=x|Y=y) \cdot P(X=x|Y=y)$$
$$= \sum_y P(Y=y) \sum_x P(X=x|Y=y)^2$$
$$\mathbb{E}_x P(X) = \sum_x P(X=x)^2$$
$$= \sum_x \left( \sum_y P(X=x|Y=y)P(Y=y) \right)^2$$

By Jensen's inequality on the square function, which is convex:
$$\left( \sum_y P(X=x|Y=y)P(Y=y) \right)^2 \leq \sum_y P(Y=y)P(X=x|Y=y)^2$$

Therefore:
$$\mathbb{E}_x P(X) \leq \mathbb{E}_{x,y} P(X|Y) \Rightarrow \frac{\mathbb{E}P(X|Y)}{\mathbb{E}P(X)} \geq 1 \Rightarrow \mathsf{I}_c(X:Y) \geq 0.$$

∎

We now prove Lemma 3.4, restated below.

**Restated Lemma** (Convexity of Collision Mutual Information): Let $p(X \mid Y) = \alpha q(X \mid Y) + (1-\alpha)r(X \mid Y)$ for $\alpha \in [0,1]$ and $p(X) = q(X) = r(X)$. Then
$$\mathsf{I}_{c_p}(X:Y) \leq \alpha \mathsf{I}_{c_q}(X:Y) + (1-\alpha)\mathsf{I}_{c_r}(X:Y).$$

*Proof*: We have the following chain of equalities:
$$\mathsf{I}_{c_p}(X:Y) = \mathsf{H}_{c_p}(X) - \mathsf{H}_{c_p}(X \mid Y)$$
$$= \frac{\mathbb{E}[\Pr[X \mid Y]]}{\mathbb{E}[\Pr[X]]}$$
$$\leq \frac{\alpha \mathbb{E}_q[\Pr[X \mid Y]] + (1-\alpha)\mathbb{E}_r[\Pr[X \mid Y]]}{\mathbb{E}[\Pr[X]]}$$

where the last equality follows from
$$\mathbb{E}[\Pr[X \mid Y]] = \sum_{x,y} p(x,y)p(y \mid x) \cdot p(x) = \sum_x p(x) \sum_y p(x \mid y)^2$$
$$= \sum_x p(x) \sum_y \alpha q(x \mid y)^2 + (1-\alpha)r(x \mid y)^2$$
$$\leq \alpha \sum_x p(x) \sum_y q(x \mid y)^2 + (1-\alpha) \sum_x p(x) \sum_y r(x \mid y)^2$$

by the convexity of $f(a) = a^2$.

∎

# B) Proofs for Facts About $2 \mapsto 1$ Quantum Random Access Code

We first prove Lemma 4.1, restated below.

**Restated Lemma** (Independence Implies Subadditivity): Given $\mathcal{E}(b_0^1, b_1^1), ..., \mathcal{E}(b_0^m, b_1^m)$, where $b_0^i, b_1^i$ for $i \in [m]$ are independently and uniformly sampled, then

$$\sup_\mu \max_{y \in \{0,1\}^m} I_c\left(b_{y_0}^1 b_{y_1}^2 ... b_{y_m}^m : \mu\left(\bigotimes_{i \in [m]} \mathcal{E}(b_0^i, b_1^i)\right)\right) \leq \sum_{i \in [m]} I_> = m \cdot I_>$$

and

$$\sup_\mu I_c\left(b_0^1 b_1^1 ... b_0^m b_1^m : \mu\left(\bigotimes_{i \in [m]} \mathcal{E}(b_0^i, b_1^i)\right)\right) \leq \sum_{i \in [m]} I_{\text{tot}} = m \cdot I_{\text{tot}}.$$

*Proof*: The proof proceeds similarly to the proof of additivity of independent states in [42] for both inequalities. Thus, we will only show the proof for $I_>$. The proof is by induction on $m$. For $m = 1$, the statement is trivially true. Then, assume that the statement is true for $m = k$. Now, consider $m = k + 1$.

$$\sup_\mu \max_{s \in \{0,1\}^n} \left[ I_{c_\mu}\left(b_{s_1}^1 b_{s_2}^2 ... b_{s_m}^m : \bigotimes_{i \in [k+1]} \mathcal{E}(b_0^i, b_1^i)\right) \right]$$

$$\leq \sup_{\mu_{1...k}, \mu_m} \max_{s \in \{0,1\}^n} \left[ I_{c_{\mu_{1...k}}}\left(b_{s_1}^1 b_{s_2}^2 ... b_{s_m}^m : \bigotimes_{i \in [k]} \mathcal{E}(b_0^i, b_1^i)\right) \right.$$

$$\left. + I_{c_{\mu_m}}\left(b_{s_{k+1}}^{k+1} : \mathcal{E}\left(b_0^{k+1}, b_1^{k+1}\right)\right) \right]$$

$$= \sum_{i \in [k]} I_> + \sup_\mu \max_{\alpha \in \{0,1\}} \left( I_c\left(b_\alpha^{k+1} : \mathcal{E}\left(b_0^{k+1}, b_1^{k+1}\right)\right) \right)$$

$$\leq \sum_{i \in [k]} I_> + I_>$$

where the first inequality follows from the chain rule of (collision) mutual information (as $\mathcal{E}\left(b_0^{k+1}, b_1^{k+1}\right)$ is independent of $\mathcal{E}(b_0^1, b_1^1), ..., \mathcal{E}(b_0^k, b_1^k)$) and by the fact that independently varying $\mu_{1...k}$ and $\mu_m$ can only increase the mutual information. Also, the last equality follows by the inductive hypothesis. ∎

We now prove Lemma 4.2, restated below except that we set $\alpha = 1$ for notational simplicity.

**Restated Lemma**: Let $b_1^1, ..., b_1^m$ be independently and uniformly sampled bits and $b_0^1, ..., b_0^m$ be arbitrary bits.

$$\sup_\mu I_c\left(b_\alpha : \mu\left(\bigotimes_i \mathcal{E}(b_0^i, b_1^i) \,\middle|\, b_{1-\alpha}^1 ... b_{1-\alpha}^m\right)\right) \leq \sum_{i \in [m]} \tilde{I}_> = m \cdot \tilde{I}_>.$$

*Proof*: Note that

$$I_c\left(b_1 : \bigotimes_i \mathcal{E}(b_0^i, b_1^i) \,\middle|\, b_0^1 ... b_0^m\right) \leq \sum_i \sup_{\mu_i} I_c(b_1^i : \mu_i(\mathcal{E}(b_0^i, b_1^i)) \,|\, b_0)$$

as given $b_0$ and the fact that $b_1^i, b_1^j$ are independent for $i \neq j$, we can have any measurement $\mu_i$ internally simulate $\mathcal{E}\left(b_0^j, b_1^j\right)$ for $j \neq i$. Then, by the definition of conditional mutual information, we have our desired result. ∎

## B.1) Proofs and Bounds for Bit Accessible Mutual Information

To obtain concrete bounds on $I_>, I_{\text{tot}}, \tilde{I}_>$, we use both analytical and computational methods.

We can use the convexity of logless mutual information and the convexity of extremal POVMs to constrain our optimization. Without loss of generality, we will prove the bound for $I_{\text{tot}}$ in this subsection. The bounds for $I_>, \tilde{I}_>$ are derived analogously. Specifically, using brute-force search over a discretized set of measurements, we will find $\mathsf{Extr}(I_{\text{tot}})$ where the POVMs, $\mu$, are over the set of extremal POVMs. Then, by convexity, we will show how to bound all other potential measurements by $\mathsf{Extr}(I_{\text{tot}})$.

We first make use of the following lemma which follows from Corollary 2.48 and Corollary 2.49 of Ref. [43].

**Lemma 2.1** (Extremal POVMs, [43]): Given POVMs $A_1, ..., A_N$ on a $D$ dimensional Hilbert space, if $N > D^2$, then there exists a decomposition of POVMs into $D^2$ POVMs $B_1, ..., B_{D^2}$ such that for all $i \in [D^2]$,
$$A_i = \sum_{j \in [N]} \alpha_{i,j} B_j$$
for probability vectors $\alpha_i$.

We then make use of convexity of collision mutual information to get the following lemma:

**Lemma 2.2**: The extremal values of $I_{\text{tot}}$ upper bound all other values of $I_{\text{tot}}$:
$$I_{\text{tot}} \leq \mathsf{Extr}(I_{\text{tot}})$$

*Proof*: The proof follows from the convexity of collision mutual information and the fact that the POVMs are a convex combination of extremal POVMs by Corollary 2.48 and Corollary 2.49 of Ref. [43]. So, for any POVM $\mu$, we can write $\mu = \sum_{i \in [\Gamma]} \alpha_i \mu_i$ for extremal POVMs $\mu_i$. Then, by convexity of collision mutual information, we have our desired result. ∎

We also state the following fact which can be seen via a straightforward use of SymPy [44] to derive the Hessian.

**Fact 2.1**: For fixed and PSD $M$ and state $\rho$,
$$f(\Delta) = \frac{\text{Tr}[(M+\Delta)\rho]^2}{\text{Tr}[M+\Delta]}$$
is convex in $\Delta$ if $M + \Delta$ is restricted to be PSD.

We can then use an $\epsilon$-net (with $\epsilon = 0.005$) to discretize the set of extremal POVMs by discretizing the possible entries of the PSD matrices and brute force search. In more detail, we make use of the following lemma:

**Lemma 2.3**: Let $M_i'$ be a PSD operator with $0 \leq M_i' \leq \mathbb{I}$ and $M_i$ be any POVM with

$$0 \leq M_i[a,b] - M_{i'}[a,b] \leq \epsilon,$$

then, for density operator $\rho$ with real entries,

$$\frac{\text{Tr}[M'_i \rho]^2}{\text{Tr}[M'_i]} \leq \frac{\text{Tr}[M_i \rho]^2}{\text{Tr}[M_i]} \leq \max_{\Delta \in \text{Extr}} \frac{\text{Tr}[(M_i + \Delta)\rho]^2}{\text{Tr}[M_i + \Delta]}$$

where $\text{Extr}$ is the set $\left\{ \begin{pmatrix} a & b \\ b & c \end{pmatrix} \,\middle|\, a,b,c \in \{0,\epsilon\} \right\}$. And also,

$$I_{\text{tot}} \leq 2 + \max_{\mu'}\left( \log_2\left( \sum_{b_0} \sum_i \max_{\Delta \in \text{Extr}} \frac{\text{Tr}\left[(M'_i + \Delta)\rho_{b_0}\right]^2}{\text{Tr}[M'_i + \Delta]} \right) + \log_2\left( \sum_{b_1} \sum_i \frac{\text{Tr}\left[(M'_i + \Delta)\rho_{b_1}\right]^2}{\text{Tr}[M'_i + \Delta]} \right) \right)$$

where $\mu'$ is the $\epsilon$-net discretization of $\mu$.

*Proof*: We first prove the second part of the claim assuming the first:

$$I_{\text{tot}} = \max_\mu \mathsf{I}_c(b_0 : \mu) + \mathsf{I}_c(b_1 : \mu) = \mathsf{H}_c(b_0) + \mathsf{H}_c(b_1) - \mathsf{H}_c(b_0 \mid \mu) - \mathsf{H}_c(b_1 \mid \mu)$$

$$= \max_\mu 2 - (\mathsf{H}_c(b_0 \mid \mu) + \mathsf{H}_c(b_1 \mid \mu))$$

$$= 2 + \max_\mu \left( \log_2\left( \sum_{b_0} \sum_i \frac{\Pr[\mu = i, b_0]^2}{\text{Tr}[\mu = i]} \right) + \log_2\left( \sum_{b_1} \sum_i \frac{\Pr[\mu = i, b_1]^2}{\text{Tr}[\mu = i]} \right) \right)$$

$$= 2 + \max_{\mu = M_1, \ldots, M_4}\left( \log_2\left( \sum_{b_0} \sum_i \frac{\text{Tr}\left[M_i \rho_{b_0}\right]^2}{\text{Tr}[M_i]} \right) + \log_2\left( \sum_{b_1} \sum_i \frac{\text{Tr}\left[M_i \rho_{b_1}\right]^2}{\text{Tr}[M_i]} \right) \right)$$

$$\leq 2 + \max_{\mu'}\left( \log_2\left( \sum_{b_0} \sum_i \max_{\Delta \in \text{Extr}} \frac{\text{Tr}\left[(M'_i + \Delta)\rho_{b_0}\right]^2}{\text{Tr}[M'_i + \Delta]} \right) + \log_2\left( \sum_{b_1} \sum_i \frac{\text{Tr}\left[(M'_i + \Delta)\rho_{b_1}\right]^2}{\text{Tr}[M'_i + \Delta]} \right) \right)$$

where the last inequality follows from the first part of the claim.

We now prove the first part of the claim. By Fact 2.1, we now that $f(\Delta) = \frac{\text{Tr}[(M+\Delta)\rho]^2}{\text{Tr}[M+\Delta]}$ is convex in $\Delta$. So, if each entry of $\Delta$ is in-between $0$ and $\epsilon$, we know that $f$ is maximized at the extremal values of $\Delta$. Moreover, note that we do not need to consider POVMs $M_i$ which have non-real entries by the following observation: take POVM element $M_i = \begin{pmatrix} a & b \\ b & c \end{pmatrix}$ for real $a,b,c$ and $\widetilde{M}_i = \begin{pmatrix} a & b+ix \\ b-ix & c \end{pmatrix}$, then

$$\text{Tr}[M_i \rho]^2 = \text{Tr}\left[\widetilde{M}_i \rho\right]^2 \quad \text{and} \quad \text{Tr}[M_i] = \text{Tr}\left[\widetilde{M}_i\right].$$

as $\rho$ has only real entries. So, given that each $M_i$ with only real entries can be expressed by $M'_i + \Delta$ for $\Delta \in \left\{ \begin{pmatrix} a & b \\ b & c \end{pmatrix} \,\middle|\, a,b,c \in [0,\epsilon] \right\}$, we have our desired result. ∎

We now have a way to upper-bound $I_{\text{tot}}$ using the $\epsilon$-net discretization of the extremal POVMs. We then perform a search[8] over the discretized set of POVMs to find the maximum value of $I_{\text{tot}}, I_>, \widetilde{I}_>$.

**Fact 2.2**: Given the $2 \mapsto 1$ QRAC scheme of Definition 2.2, we use brute force search over an $\epsilon$-net to find that

$$I_> \leq 0.58$$
$$\tilde{I}_> \leq 0.58$$
$$I_{\text{tot}} \leq 0.67.$$

---

[8]Technically, our code-base does not quite employ brute-force search as we employ a "progressive" refinement approach to speedup the search. Specifically, we first search over a coarse grid and then refine the grid around the maximum value. At each step of the refinement, we "throw away" the points which provably cannot be the maximum value.

## C Existance of Parameter Bounds

We show that the parameters in Section 6 are well-defined.

**Lemma 3.1** (Existence of Parameters): For any fixed $\epsilon', \epsilon'', d, D$, there exists a $n$ and $r$ such that

$$2\log\left(\frac{1}{\epsilon'}\right) + \log\left(\frac{1}{\epsilon''}\right) + 4\,|\overline{\text{CU}}| + (2r)^D \leq \frac{n}{100}$$

and

$$|\overline{\text{CU}}| \geq \log\left(\frac{1}{\epsilon'}\right).$$

*Proof*: Note that $2\log(\frac{1}{\epsilon'}) + \log(\frac{1}{\epsilon''})$ are constants and so, we just have to bound the term $2\,|\overline{\text{CU}}| + (2r)^D$. Then,

$$|\overline{\text{CU}}| = n \cdot \left(1 - \frac{(2r)^D}{(2r + 2\ell^d)^D}\right)$$

and so

$$4|\overline{\text{CU}}| + (2r)^D \leq 4n \cdot \left(\left(1 - \frac{(2r)^D - \frac{(2r \cdot (2r + 2\ell^d))^D}{n}}{(2r + 2\ell^d)^D}\right)\right).$$

Define

$$E = \frac{(2r)^D - \frac{(2r \cdot (2r + 2\ell^d))^D}{n}}{(2r + 2\ell^d)^D}.$$

Then, to satisfy the inequality, we need to show the exisrance of some growth function for $n$ and $r$ such that $E$ approaches 1. Clearly, we can arbitrarily set $r$ such that for all $\gamma$,

$$E = \frac{(2r)^D}{(2r + 2\ell^d)^D} \geq 1 - \gamma.$$

We can then set $n$ such for all $\gamma'$ and $r$

$$E = \frac{\left(2r \cdot \left(2r + 2\ell^d\right)\right)^D}{n} < \gamma'.$$

Finally, note that as $\log(\frac{1}{\epsilon'})$ is a constant, we can always find $n$ such that

$$|\overline{\text{CU}}| \geq \log\left(\frac{1}{\epsilon'}\right),$$

completing the proof.

∎